\documentclass[aps,superscriptaddress]{revtex4-2}
\usepackage{amssymb,amsmath,epsfig}
\usepackage{subfigure}
\usepackage{color}
\usepackage[colorlinks=true, pdfstartview=FitV, linkcolor=blue, citecolor=red, urlcolor=magenta, breaklinks=true]{hyperref}

\begin{document}
\title{Scattering and absorption by extra-dimensional black holes with GUP}

\author{M. A. Anacleto}\email{anacleto@df.ufcg.edu.br}
\affiliation{Departamento de F\'{\i}sica, Universidade Federal de Campina Grande
Caixa Postal 10071, 58429-900 Campina Grande, Para\'{\i}ba, Brazil}

\author{J. A. V. Campos}\email{joseandrecampos@gmail.com}
\affiliation{Departamento de F\'isica, Universidade Federal da Para\'iba, 
Caixa Postal 5008, 58051-970 Jo\~ao Pessoa, Para\'iba, Brazil}

\author{F. A. Brito}\email{fabrito@df.ufcg.edu.br}
\affiliation{Departamento de F\'{\i}sica, Universidade Federal de Campina Grande
Caixa Postal 10071, 58429-900 Campina Grande, Para\'{\i}ba, Brazil}
\affiliation{Departamento de F\'isica, Universidade Federal da Para\'iba, 
Caixa Postal 5008, 58051-970 Jo\~ao Pessoa, Para\'iba, Brazil}

\author{E. Maciel}\email{eugeniobmaciel@gmail.com}
\affiliation{Departamento de F\'{\i}sica, Universidade Federal de Campina Grande
Caixa Postal 10071, 58429-900 Campina Grande, Para\'{\i}ba, Brazil}
 
\author{E. Passos}\email{passos@df.ufcg.edu.br}
\affiliation{Departamento de F\'{\i}sica, Universidade Federal de Campina Grande
Caixa Postal 10071, 58429-900 Campina Grande, Para\'{\i}ba, Brazil}

\begin{abstract}  

In this paper, we consider the Schwarzschild-Tangherlini black hole to investigate the process of scalar wave scattering by the black hole in a spacetime of $(d + 1)$ dimensions and also with the generalized uncertainty principle (GUP). In this scenario, we analytically determine the phase shift and explore the effect of extra dimensions by calculating the differential scattering and absorption cross-section by applying the partial wave method at low and high-frequency limits. We show at high dimensions that the absorption is not zero as the mass parameter approaches zero. 

\end{abstract}

\maketitle
\pretolerance10000

\section{Introduction}

Black holes are mysterious objects in our universe and are one of the most important predictions of general relativity. Furthermore, great advances in exploring the physics of black holes have been made for many decades. Thus, for a better understanding of black hole physics, several black hole solutions were constructed. The study of black holes works as a large laboratory in the perspective of analyzing gravitational radiation and its oscillations. In addition, by exploiting this radiation, we can obtain internal information about the characteristics of the black hole, for example, its mass and charge. It is worth mentioning that, recently, the LIGO-VIRGO collaboration~\cite{LIGOScientific:2016aoc,LIGOScientific:2017vwq} presented the result of the detection of gravitational waves arising from the merger of a binary black hole system. Also, in recent years, the Event Horizon Telescope, with remarkable observation, showed the image of the shadow of the massive black hole at the center of the galaxy M87~\cite{EventHorizonTelescope:2019dse,EventHorizonTelescope:2019ggy}. Hence, these observations point to the existence of black holes. And thus, new directions of research arise in general relativity, modified theories of gravity, and relativistic astrophysics. Thus, leading to a great depth in the development of black hole physics. In particular, interest in studying the scattering process in gravitational theories has increased in recent years due to such observations. In this paper we consider the effect of extra dimensions to examine the process of scattering scalar waves by a Schwarzschild-Tangherlini black hole~\cite{Tangherlini:1963bw} that is a solution of Einstein's field equations in a spacetime of $(d+1)$ dimensions.

The physics of black holes in extradimensional spacetimes has gained increasing interest in recent years. Moreover, it is well known that higher dimensional spacetimes contain more physical information. Since the black hole is not an isolated system, it interacts with its surroundings through Hawking radiation, scattering, and absorption. Hawking radiation from the Schwarzschild-Tangherlini black hole has been investigated in~\cite{Feng:2015jlj}. Thus, due to these interactions, we can explore their properties by computing the process of scattering and absorption of matter waves by black holes. Interest in the study of scattering by black holes has been reported by Matzner~\cite{Matzner1968}. Furthermore, this has stimulated several studies to investigate many aspects of the process of scattering and absorption of scalar waves by black holes~\cite{Futterman1988,Matzner:1977dn,Westervelt:1971pm,Peters:1976jx,Sanchez:1976fcl,Sanchez:1976xm,DeLogi:1977dp,Doran:2001ag,Dolan:2007ut,Crispino:2009ki,Churilov1974,Gibbons:1975kk,Page:1976df,Churilov1973,Jung:2004yh,Doran2005,Dolanprd2006,Castineiras2007,Benone:2014qaa,Das:1996we,deOliveira:2018kcq,Hai:2013ara}. Besides, initial work on the numerical analysis of the scattering of matter waves by black holes has been carried out by S\'anchez~\cite{Sanchez1978,NSanchez1978}. On the other hand, considering modified gravity models, the process of scattering matter waves by black holes has also been examined for various types of metrics~\cite{Campos:2021sff,Anacleto:2021qoe,Anacleto:2018acl,Li:2022wzi,Xing:2022emg,Bisnovatyi-Kogan:2022ujt,Zeng:2021dlj,Mourad:2021qgo,Jha:2021bue,Li:2022jda,Gogoi:2022wyv,Karmakar:2022idu,Lobos:2022jsz,Tsupko:2022yzg,Zeng:2022fdm,Heydari-Fard:2021qdc,Heydari-Fard:2021pjc,Khodadi:2021gbc,Fathi:2020sfw,Chen:2022ngd,Filho:2023etf,Heidari:2023bww}. In addition, the analysis of the scattering of matter waves by black holes in high dimensions has been reported in~\cite{Myers:1986un,Jung:2004yn,Moura:2011rr,Marinho:2016ixt,Ahmedov:2021ohg,Singh:2017vfr,Moura:2021eln}.
In~\cite{Tsukamoto:2014dta} the authors have studied the effects of gravitational lensing on the strong field limit in Tangherlini spacetime.

General relativity predicts the existence of singularities. In particular, black hole solutions are often affected by singularities. However, these singularities are usually expected to be found in the inner region of a black hole even for extra dimensions. In recent years, when considering the generalized uncertainty principle (GUP)~\cite{das2008universality,das2009phenomenological,ali2011proposal,buoninfante2020phenomenology,Lake:2018zeg,Lake:2019nmn,Casadio:2020rsj,Buoninfante:2020guu}, these quantum corrections of black hole event horizons imply the formation of new types of horizonless objects, i.e., the effective radius is greater than the horizon radius of Schwarzschild. Furthermore, in~\cite{Anacleto:2020lel}, we consider the Schwarzschild black hole with quantum corrections implemented by GUP to calculate the absorption and differential scattering cross-section. In this work, we will verify if this effect remains or not when we consider the effect of quantum corrections due to GUP in the scattering process for a Schwarzschild-Tangherlini black hole. 
Our purpose in this work is to examine the process of scattering scalar waves by a Schwarzschild-Tangherlini black hole in a spacetime of $(d + 1)$ dimensions by considering quantum corrections due to GUP. 
In this approach, the effect of extra dimensions will be incorporated into the modified dispersion relation obtained from the generalized uncertainty principle~\cite{Anacleto:2021qoe,Anacleto:2020lel} (see also~\cite{Koppel:2017rsf,Carr:2022ndy,Scardigli:2008jn,Scardigli:2003kr}).
Therefore, we will apply the partial wave approach to determine the differential scattering and absorption cross-sections in the low and high-frequency regimes. Furthermore, we will also calculate such quantities via the numerical method, which is valid for arbitrary values of frequencies. We will also determine the differential scattering/absorption cross-section in the high-frequency regime by the null geodesic method. Thus, we will adopt the technique introduced in the works~\cite{Anacleto:2022shk,Anacleto:2017kmg,Anacleto:2019tdj,Anacleto:2020zhp,Anacleto:2020lel}.

The paper is organized as follows. In Sec.~\ref{s1} we introduce the Schwarzschild-Tangherlini black hole considering quantum corrections due to GUP. In Sec~\ref{s2} we derive the phase shift and calculate the differential scattering/absorption cross section for a Schwarzschild-Tangherlini black hole by considering analytical and numerical analysis. 
In Sec.~\ref{na} we present the numerical analysis.
In Sec.~\ref{conc} we make our final considerations.

\section{Black Hole in extra space dimensions}
\label{s1}
The Tangherlini spacetime or more precisely the Schwarzschild-Tangherlini spacetime~\cite{Tangherlini:1963bw} consists of a solution of Einstein's field equations in a spacetime of $(d+1)$ dimensions. Thus, the external solution for a specifically symmetric and neutral black hole in extra dimensions spacetime is described by the metric
\begin{equation}
ds^2 = \left(1-\dfrac{2\mu}{r^{d-2}}\right)dt^2 - \left(1-\dfrac{2\mu}{r^{d-2}}\right)^{-1}dr^2 - r^2 d\Omega_{d-1}^2,
\label{metric1}
\end{equation}
where 
\begin{eqnarray}
d\Omega_{d-1}^2 = d\theta_{d-2}^2 + \sin^2{\theta_{d-2}^2}\left[d\theta_{d-3}^2 + \sin^2\theta_{d-3}\left(... + \sin^2\theta_{2}\left(d\theta_{1}^2+\sin^2\theta_{1}d\phi^2\right)...\right) \right],
\end{eqnarray}
is the solid angle for the space of $(d-1)$ dimensions and $\Omega_{d-1}$ is the volume of a sphere $(d-1)$ dimensions $\Omega_{d-1} = {2\pi^{d/2}}/{\Gamma(d/2)}$.

The parameter $\mu$ is one constant and is related to the black hole mass $M$ as follows:
\begin{equation}
\mu = \dfrac{16\pi M}{2(d - 1)\Omega_{d-1}}=\dfrac{8 M\Gamma(d/2)}{2(d-1)\pi^{(d-2)/2}},
\end{equation}
Note that by setting $d=3$ in \eqref{metric1} the metric returns the ordinary Schwarzschild spacetime. 
In addition, if the mass parameter $\mu$ is less than zero we will have a non-physical situation. However, for this parameter greater than zero, the radius of the horizon can be obtained by taking $g_{tt}(r_{h})=0$
\begin{equation}
\label{rh}
r_{h}^{d-2}=2\mu = \dfrac{16\pi M}{(d-1)\Omega_{d-1}}
=\dfrac{8 M\Gamma(d/2)}{(d-1)\pi^{(d-2)/2}}.
\end{equation}
For $d=3$, we recover the radius of the event horizon of the usual Schwarzschild black hole, $r_h=2M$.  
We can also write the black hole mass, $M$, in terms of the event horizon as follows
\begin{eqnarray}
M = \left[\dfrac{(d-1)\Omega_{d-1}}{16\pi}\right]r_{h}^{d-2}
=\left[\dfrac{8 M\Gamma(d/2)}{(d-1)\pi^{(d-2)/2}}\right]r_{h}^{d-2}.
\end{eqnarray}
In this way, we can still define the event horizon area in metric \eqref{metric1} 
\begin{eqnarray}
A = \Omega_{d-1}r_{h}^{2(d-2)}=\dfrac{2\pi^{d/2}}{\Gamma(d/2)}r_{h}^{2(d-2)}.   
\end{eqnarray}
In this case, from metric \eqref{metric1} we find that the Hawking temperature of the Schwarzschild–Tangherlini is given by
\begin{eqnarray}
 T_{Hst}=\frac{d-2}{4\pi r_h}.  
\end{eqnarray}
Hence, for $d=3$ the Hawking temperature of the Schwarzschild black hole is recovered. 
In Fig.~\ref{figTemp}, the behavior of the Hawking temperature is shown when we vary the dimension parameter $d$.

\subsection{Quantum-corrections to the metric}
{
At this point, we will now apply quantum corrections to the metric of a Schwarzschild-Tangherlini black hole and thus investigate the influence of these corrections on the scattering process. 
In this way, we introduce the effect of the extra dimensions into the modified dispersion relation that has been derived from the GUP~\cite{Anacleto:2020lel,Anacleto:2021qoe}.
As shown in \cite{Anacleto:2020lel} and recently applied in studies on quasinormal modes~\cite{Anacleto:2021qoe}, the metric can be constructed by rewriting the horizon radius. So let us consider the following definition} \cite{medved2004conceptual, carr2015sub, tawfik2014generalized, kempf1995hilbert,Pedram:2011gw}
\begin{eqnarray}
\Delta x\Delta p \geq \dfrac{\hbar}{2}\left[1 - \dfrac{\alpha l_{p}}{\hbar} \Delta p + \dfrac{\beta l_{p}^2}{\hbar^2} (\Delta p)^2\right],
\label{appGUP}
\end{eqnarray}
where $\alpha$ and $\beta$ are dimensionless positive parameters and $l_{p}$ is the Planck length. 
We shall here adopt without loss of generality, the natural units $G = c = k_{B} = \hbar = l_{p} = 1 $.
In this way, we assume that $\Delta p \sim p \sim E$ and from the equation \eqref{appGUP} 
we obtain~\cite{Anacleto:2021qoe,Anacleto:2020lel}
\begin{eqnarray}
\mathcal{E}  \geq  E\left[1 - \dfrac{4\alpha}{\Delta x} + \dfrac{16\beta}{(\Delta x)^2} + \cdots \right] 
= E\left[1 - \dfrac{4\alpha}{r_{h}} + \dfrac{16\beta}{r_{h}^{2}} + \cdots \right],
\end{eqnarray}
where we identify $\mathcal{E}$ as the energy of the black hole corresponding to the GUP and consider $\Delta x$ as the radius of the horizon. Now, we introduce the contribution of extra dimensions into the above equation by making 
$\mathcal{E} \rightarrow \mathcal{E}^{d-2}$, $E \rightarrow E^{d-2}$, and $r_h \rightarrow r_h^{d-2}$, so we have
\begin{eqnarray}
\mathcal{E}^{d-2} & \geq & E^{d-2}\left[1 - \dfrac{4\alpha}{r_{h}^{d-2}} + \dfrac{16\beta}{r_{h}^{2(d-2)}} + \cdots \right].
\label{ener}
\end{eqnarray}
Equation \eqref{ener} can be written in terms of the mass parameter by taking $E^{d-2} \sim \mu=r_{h}^{d-2}/2$ 
and $\mathcal{E}^{d-2} \sim M_{gup}^{d-2} $. 
Hence, we find the following relation
\begin{eqnarray}
M_{gup}^{d-2} \geq \frac{r_{h}^{d-2}}{2}\left[1 - \dfrac{4\alpha}{r_{h}^{d-2}} + \dfrac{16\beta}{r_{h}^{2(d-2)}} \right] 
= \mu\left[1 - \dfrac{4\alpha}{2\mu} + \dfrac{16\beta}{4\mu^{2}} \right].
\end{eqnarray}
Here $M_{gup}^{d-2}$ being the mass corresponding to the Schwarzschild-Tangherlini black hole with GUP. 

Now, for the radius of the horizon, we obtain
\begin{eqnarray}
r_{hgup}^{d-2}=2M_{gup}^{d-2}=r_{h}^{d-2}\left[1 - \dfrac{4\alpha}{r_{h}^{d-2}} + \dfrac{16\beta}{r_{h}^{2(d-2)}} \right],
\end{eqnarray}
and in terms of the mass $M$, we have
\begin{eqnarray}
r_{hgup}^{d-2}=\dfrac{8M\Gamma(d/2)}{(d-1)\pi^{(d-2)/2}}\left[1 - \dfrac{(d-1)\pi^{(d-2)/2}}{2\Gamma(d/2)} \dfrac{\alpha}{M} + \dfrac{(d-1)^{2}\pi^{(d-2)}}{4\Gamma(d/2)^{2}}\dfrac{\beta}{M^{2}} \right].
\end{eqnarray}
For $d=3$, we recover the result found in~\cite{Anacleto:2020lel,Anacleto:2021qoe}, that is
\begin{eqnarray}
r_{hgup}=r_{h}\left[1 - \dfrac{4\alpha}{r_{h}} + \dfrac{16\beta}{r_{h}^{2}} \right]
=2M\left[1 - \dfrac{4\alpha}{2M} + \dfrac{16\beta}{4M^{2}} \right].
\end{eqnarray}

\subsection{Temperature Analysis}
\begin{figure}[!htb]
 \centering
\includegraphics[scale=0.3]{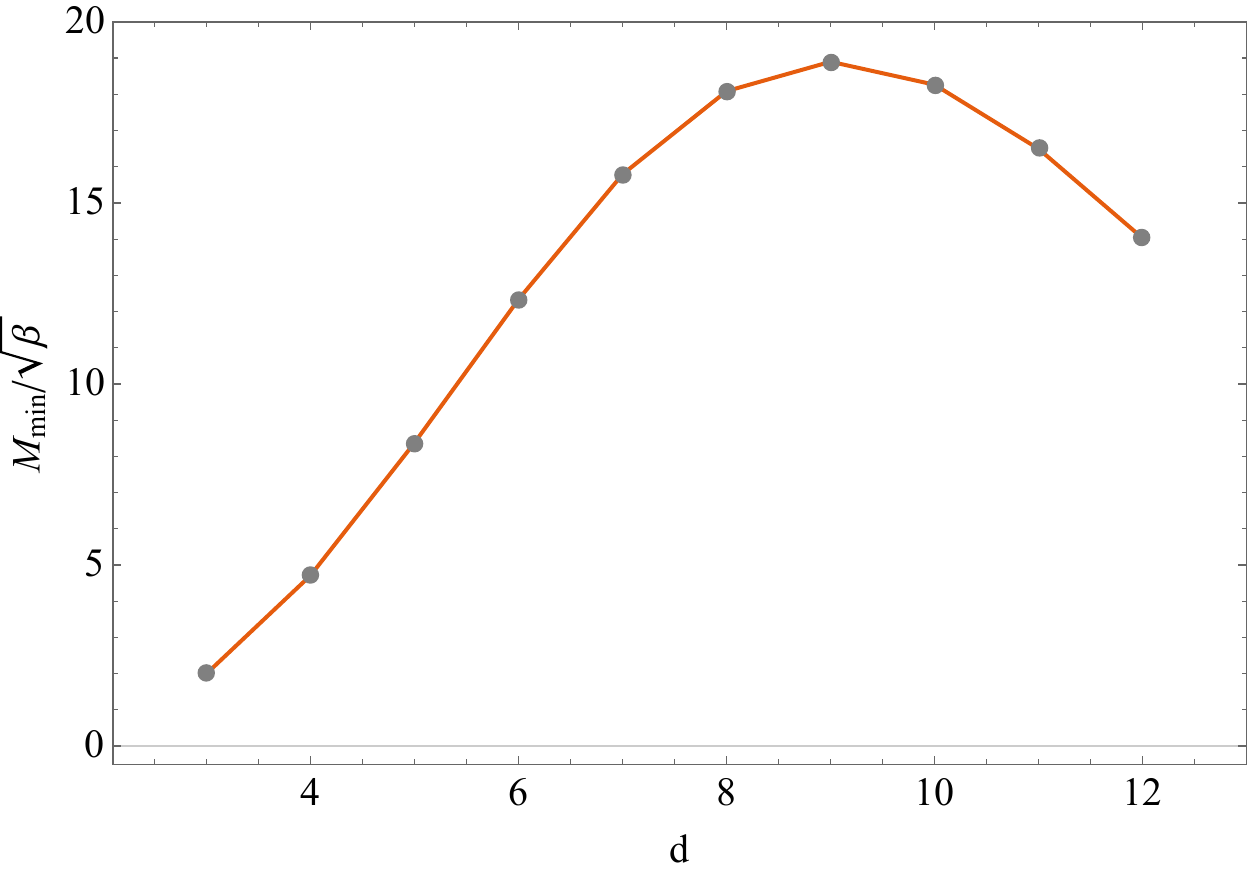}
 \\
  \caption{\footnotesize{Minimum mass behavior $M_{min}$ as function of dimension. At $d=9$ we have the highest value for $M_{min}$.}}
 \label{Mmin}
\end{figure}

\begin{figure}[!htb]
 \centering
 \subfigure[]{\includegraphics[scale=0.3]{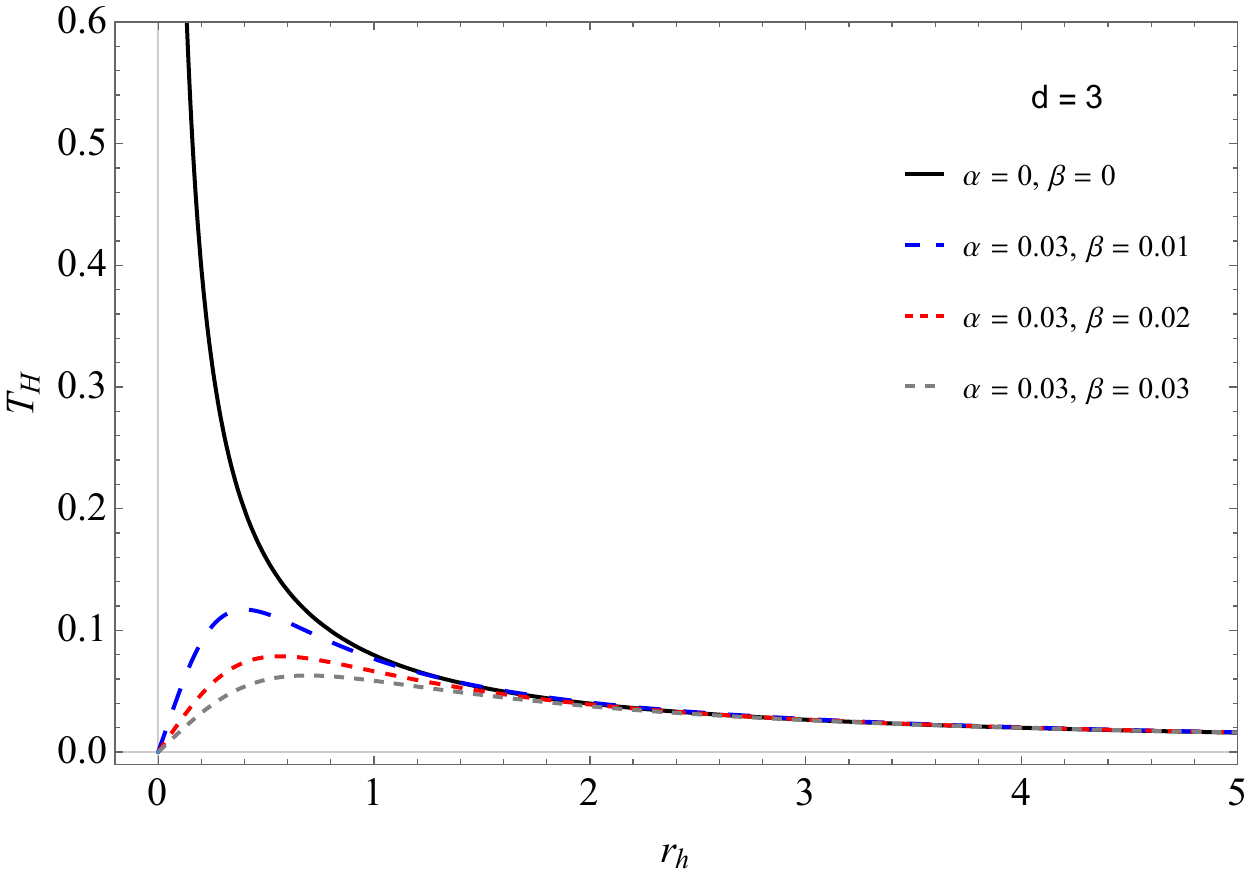}\label{Temd3}}
 \qquad
 \subfigure[]{\includegraphics[scale=0.3]{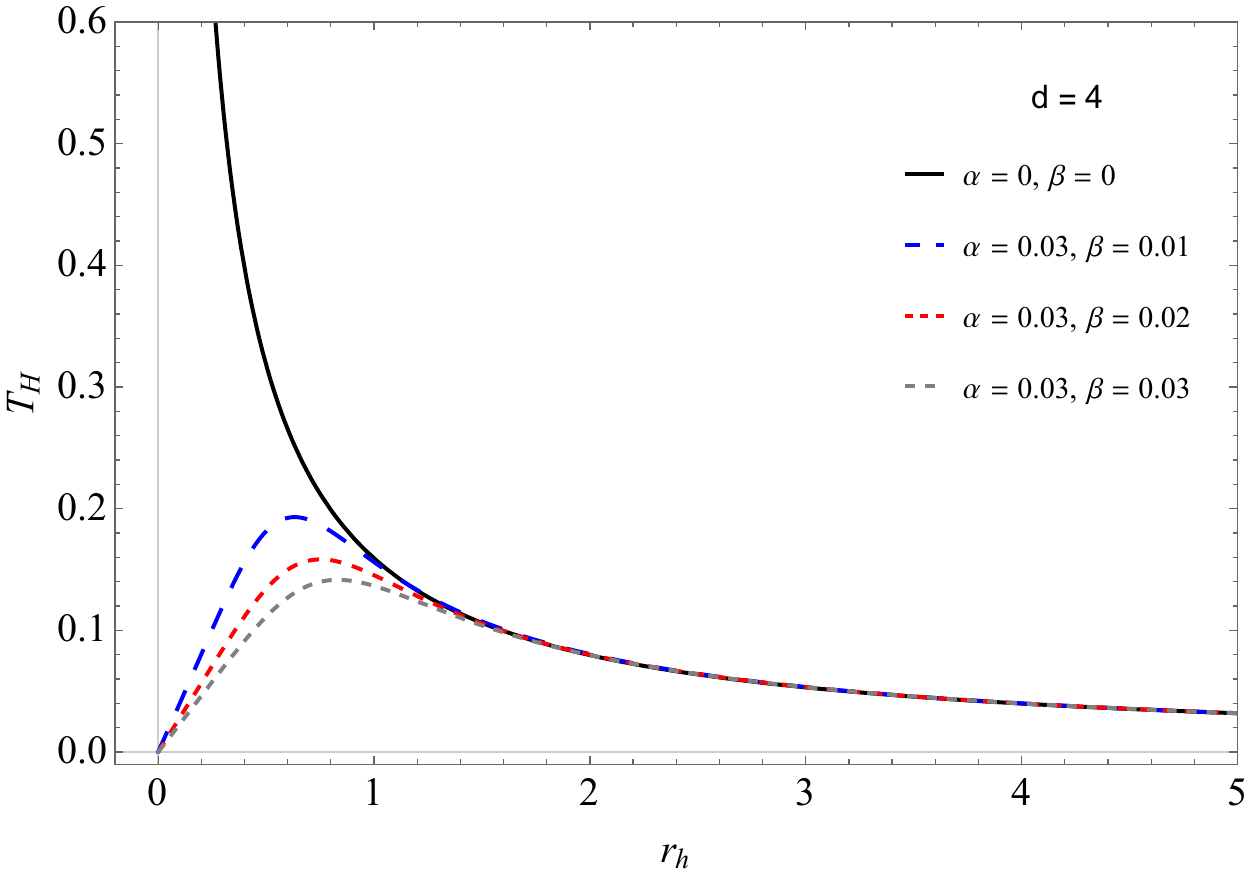}\label{Temd4}}
 \qquad
 \subfigure[]{\includegraphics[scale=0.3]{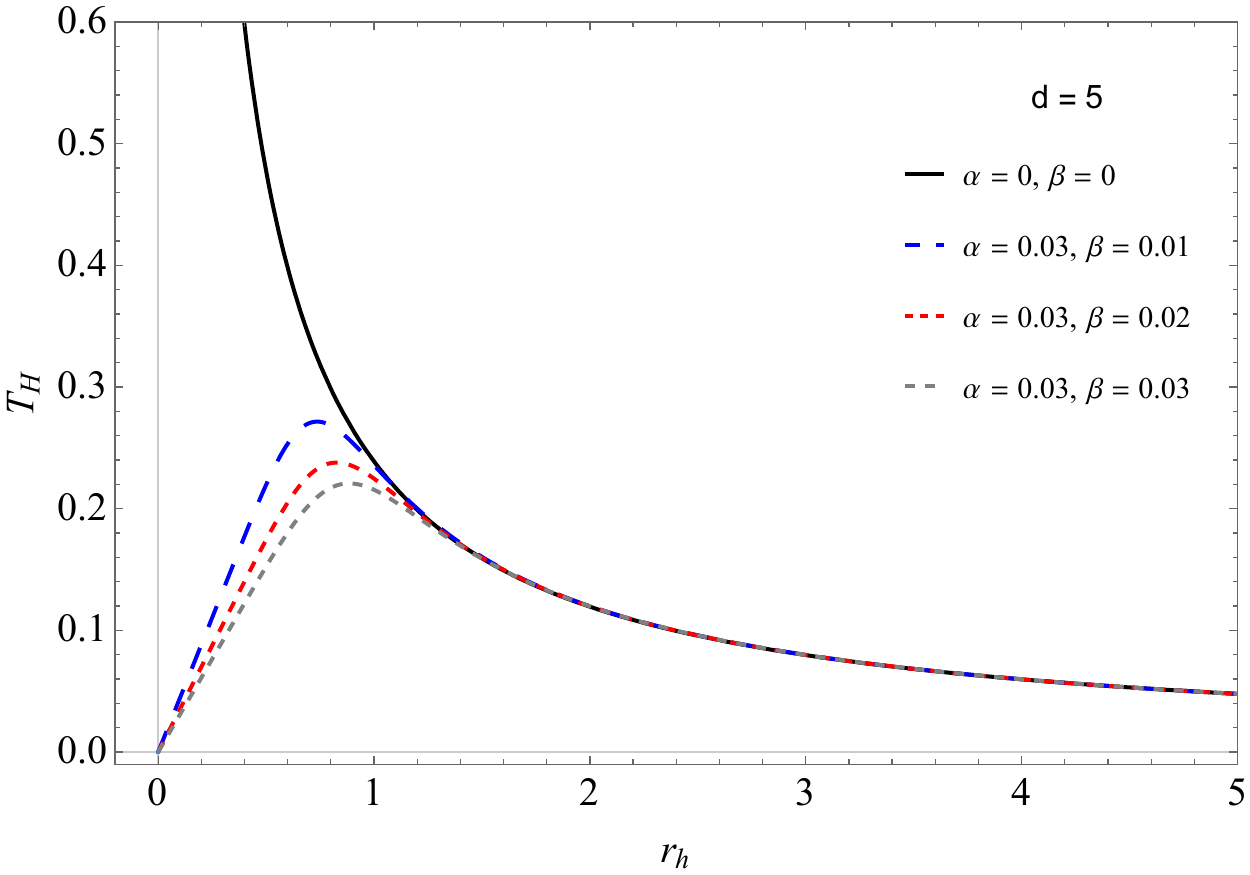}\label{Temd5}}
 \qquad
 \subfigure[]{\includegraphics[scale=0.3]{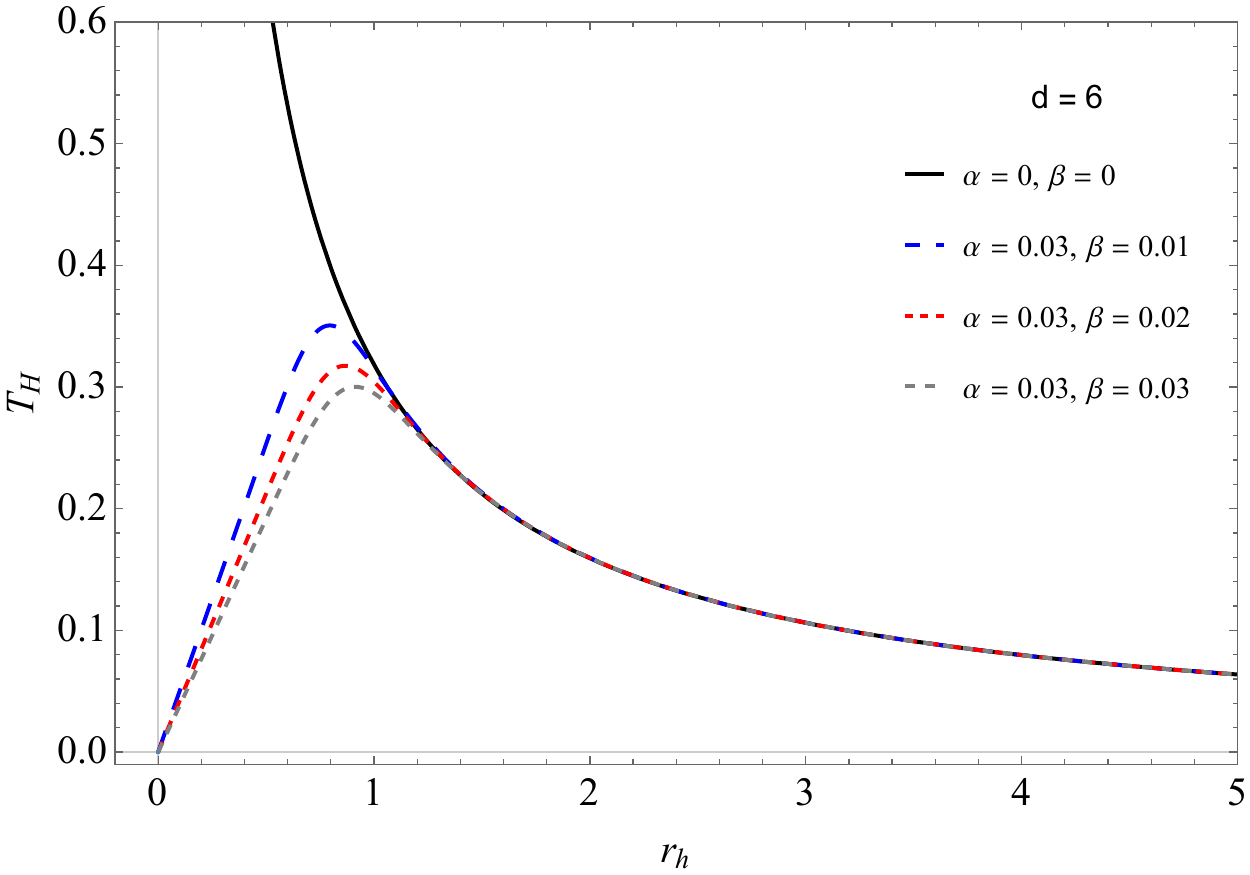}\label{Temd6}}
 \\
  \caption{\footnotesize{Graphs for the Hawking temperature as a function of the event horizon $r_{h}$. We can see the effect of the GUP corrections for each dimension.}}
 \label{figTemp}
\end{figure}

{The thermodynamic study for quantum corrections of GUP black holes has been studied in recent years~\cite{medved2004conceptual, park2008generalized,Anacleto:2015mma,gangopadhyay2015constraints, feng2016quantum, Anacleto:2021nhm,Segreto:2022clx,Feng:2016tyt,Feng:2020ams,Feng:2022gdz,Anacleto:2022sim,Scardigli:1999jh,Scardigli:2014qka,Scardigli:2016pjs,Scardigli:2018jlm,Lambiase:2017adh}. The GUP effect leads to the appearance of remnants in the final stages of black hole evaporation, as discussed in \cite{adler2001generalized,Scardigli:2010gm}.}
Here, we incorporate the effect of GUP into metric \eqref{metric1} 
by making $r_h^{d-2}=2\mu\rightarrow r_{hgup}^{d-2}$~\cite{Anacleto:2020lel,Anacleto:2021qoe}, such that the line element is now given by 
\begin{equation}
ds^2 = B(r)dt^2 - B(r)^{-1}dr^2 - r^2 d\Omega_{d-1}^2,
\label{metric2}
\end{equation}
being 
\begin{eqnarray}
B(r) = 1 - \left(\dfrac{r_{hgup}}{r}\right)^{d-2},
\end{eqnarray}
the metric function.
The Hawking temperature with GUP corrections in terms of $r_{h}$ is given by
\begin{eqnarray}
T_{Hgup}=\dfrac{(d-2)}{4\pi r_{hgup}}
=\dfrac{(d-2)}{4\pi r_{h}}\left(1 - \dfrac{4\alpha}{r_{h}^{d-2}} + \dfrac{16\beta}{r_{h}^{2(d-2)}}\right)^{-1/(d-2)}.
\label{Temp}
\end{eqnarray}
By taking $d=3$, we have $T_{Hgup}=1/4\pi r_{hgup}$, with $r_{hgup}=r_h\left(1 - {4\alpha}/{r_{h}} 
+ {16\beta}/{r_{h}^{2}}\right)$~\cite{Anacleto:2020lel}.
{In Fig.~\ref{figTemp}, we show the temperature behavior in terms of the horizon radius $r_{h}$. Note that in all scenarios, the GUP correction parameters eliminate the temperature divergence (black curve) for small values of $r_{h}$. Thus, the maximum temperature value for the GUP case is proportional to a radius $r_{h} = (16\beta)^{1/2(d-2)}$. In this way, we have a maximum temperature}
\begin{eqnarray}
T_{Hgup}^{max} = \dfrac{(d-2)}{\pi (16)^{(d-1)/2(d-2)}}\left(2\sqrt{\beta} - \alpha \right)^{-1/(d-2)}.
\end{eqnarray}
{Now, by using the equation \eqref{rh}, we have a minimum mass corresponding to a maximum temperature, for each dimension }
\begin{eqnarray}
M_{min} = \dfrac{(d-1)\pi^{(d-2)/2}}{2\Gamma(d/2)}\sqrt{\beta}.
\end{eqnarray} 
{Note that this minimum mass has an increasing behavior depending on the size up to $d = 9$, as we can see in Fig.~\ref{Mmin}. In the dimension range until $d=9$, it is interesting to note that all values have a mass value greater than one. If this is the minimum value, we have for larger dimensions a much larger mass compared to the usual dimension $d = 3$. In this way, we have that $M_{min}^{(d=9)} = 9.449 M_{min}^{(d=3)}$.}

\section{Scattering and Absorption in Extra Dimensions with GUP}
\label{s2}
In this section, we will investigate the Schwarszchild-Tangherlini black hole in space with extra dimensions and determine the differential cross section for this model. Here, we will apply the partial wave method to calculate the phase shift for a space with $ d $ spatial dimensions. For that, we will consider the case of a massless scalar field and its equation of motion to describe the wave scattering in the background field. We will also determine the absorption cross section in the extra dimensions scenario.

\subsection{Differential scattering cross section}
\begin{figure}[!htb]
 \centering
 \subfigure[]{\includegraphics[scale=0.35]{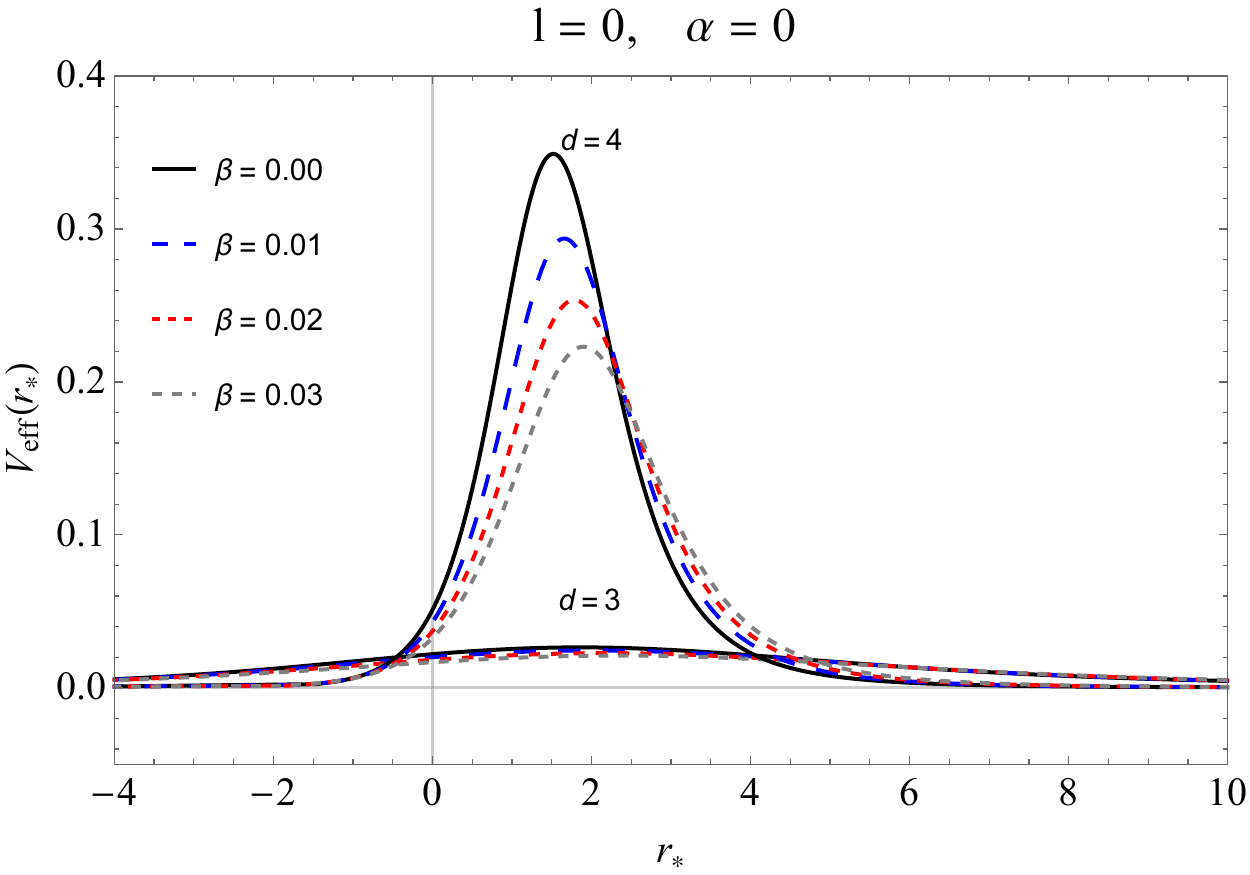}\label{potl0d3d4}}
 \qquad
 \subfigure[]{\includegraphics[scale=0.35]{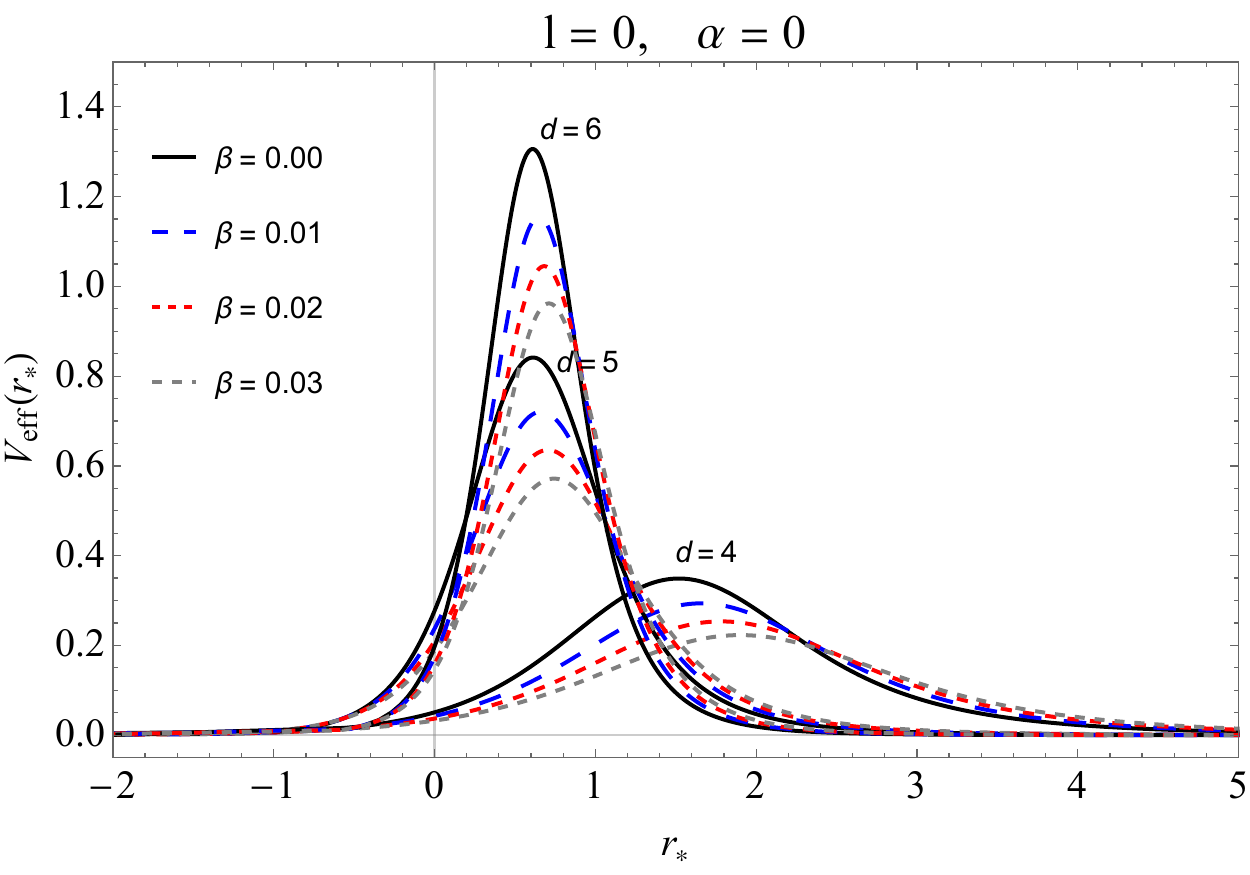}\label{potl0d4d5d6}}
 \\
  \caption{\footnotesize{The effective potential in terms of tortoise coordinates for $M=1$, $l = 0$ and with only the quadratic part of the GUP, i.e., $\alpha = 0$. In figure \ref{potl0d3d4} we compare dimensions $d = 3$ and $d = 4$, we note that the influence of corrections is greater for high dimensions. In \ref{potl0d4d5d6} the behavior of the potential for some values of extra dimensions.}}
 \label{pottot}
\end{figure}

The dynamics of a scalar field is described by the Klein-Gordon equation 
\begin{eqnarray}
\dfrac{1}{\sqrt{-g}}\partial_{\mu}\Big(\sqrt{-g}g^{\mu\nu}\partial_{\nu}\Psi\Big)=0 .
\end{eqnarray}
Here we emphasize that the contribution of GUP in the above equation will appear explicitly due to the background metric $\eqref{metric2}$. Thus, By considering the background field described by the metric $\eqref{metric2}$ and using the method of separating different variables, we can define an {\it Ansatz} of the type
\begin{eqnarray}
\Psi_{\omega l m}({\bf r},t)=\frac{{\cal R}_{\omega l}(r)}{r}Y_{lm}(\theta_{1},\phi)e^{-i\omega t},
\end{eqnarray}
with $ \omega $ the frequency an $Y_{lm}(\theta_{1},\phi)$ the spherical harmonics. In this way, we can find the following equation in terms of $ {\cal R}_{\omega l}(r) $  
\begin{eqnarray}
\label{eqrad}
B(r)\dfrac{d}{dr}\left[B(r)\dfrac{d{\cal R}_{\omega l}(r)}{dr} \right] +\left[ \omega^2 -V_{eff} \right]{\cal R}_{\omega l}(r)=0,
\end{eqnarray}
where the term $ V_{eff} $ is the effective potential defined as
\begin{eqnarray}
V_{eff}=\frac{B(r)}{r}\left[(d-2)\left(\dfrac{r_{hgup}}{r}\right)^{d-2}
+ \frac{l(l+1)}{r}\right].
\end{eqnarray}
We can write the equation \eqref{eqrad}  as a Schr\"odinger-type equation. In fact, by taking a new variable substitution $d/dr_{*} = B(r)d/dr$, well-known as tortoise coordinate, we find the equation
\begin{eqnarray}
\dfrac{d^2{\cal R}_{\omega l}}{dr_{*}^2}  +\left[ \omega^2 - V_{eff}(r_{*}) \right]{\cal R}_{\omega l}=0.
\label{eqRtot}
\end{eqnarray}
The tortoise coordinate can be determined in terms of a hypergeometric function as follows.
\begin{equation}
r_{*} = r - rF\left(1, \dfrac{1}{d-2}; \dfrac{d-1}{d-2};\left(\dfrac{r}{r_{hgup}}\right)^{d-2}\right).
\end{equation}
The graphs below show us the effective potential in terms of the tortoise coordinate $r_{*}$. Through these it is possible to observe the limits of this new coordinate. {The behavior of the potential as a function of the tortoise coordinate is shown in Fig.~\ref{pottot}. In Fig.~\ref{potl0d3d4} we have dimensions $d = 3$ and $d = 4$ where it is possible to see the scale difference between dimension $d = 3$ and the others. Moreover, the influence of the corrections coming from the GUP is much stronger on the extra dimensions. In Fig.~\ref{potl0d4d5d6} we see that, the effective potential grows in terms of the number of dimensions.}

Note from the graphs in Fig.~\ref{pottot} that the effective potential $V_{eff}(r_{*})$ goes to zero for large distances, that is, at the extremes. Also note that the extra dimension does not influence the behavior of the potential in this region. In fact, for distances greater than the length of the extra dimensions its effects should be ignored. However, if we consider regions very close to the event horizon, we can observe from the graphs that the extra dimensions significantly increase the effective potential. We can thus investigate the behavior of the function ${\cal R}_{\omega l}$ asymptotically. 
For $r_{*} \rightarrow -\infty $, $V_{eff}(r_{*}) \rightarrow 0$
\begin{equation}
{\cal R}_{\omega l} \approx C_{tr}e^{-i\omega r_{*}}.
\label{solRrh}
\end{equation}
For $r_{*} \rightarrow +\infty $, $V_{eff}(r_{*}) \rightarrow 0$,
\begin{equation}
{\cal R}_{\omega l} \approx C_{in}e^{-i\omega r_{*}} + C_{ref}e^{i\omega r_{*}},
\label{solRinf}
\end{equation}
where $C_{ref}, C_{tr}, C_{in}$ are complex constants, which satisfy the relationship $|C_{in}|^2 = |C_{ref}|^2 + |C_{tr}|^2$.

In the region $r >> r_{h}$ we can still express the radial equation in a simpler way since in this regime $r_{*} \sim r$ such that the effective potential reads 
\begin{eqnarray}
V_{eff} \approx \frac{l(l + 1)}{r_{*}^2}.
\end{eqnarray}
Thus, the solution
of the radial equation \eqref{eqRtot} can be expressed in terms of Hankel spherical functions of the first and second kind
\begin{eqnarray}
{\cal R}_{\omega l} \approx \omega r_{*}\left[ (-i)^{l+1} C_{in}H_{l}^{(2)}(\omega r_{*}) + i^{l+1} C_{ref}H_{l}^{(1)}(\omega r_{*})\right].
\end{eqnarray}
In the region $\omega r_{*} >> l(l+1)/2$,  by knowing that 
\begin{eqnarray}
&&H_{l}^{(1)} \approx \frac{(-i)^{l+1} e^{i \omega r_{*}} }{\omega r_{*}},
\\
&&H_{l}^{(2)} \approx \frac{i^{l+1} e^{-i \omega r_{*}} }{\omega r_{*} },
\end{eqnarray}
we recover \eqref{solRinf}.
The phase shift $\delta_{l}$ is defined by
\begin{eqnarray}
e^{2i\delta_{l}} = (-1)^{l+1}\dfrac{C_{ref}}{C_{in}}.
\end{eqnarray}
By considering the new radial function $ \chi(r)=\sqrt{B(r)}{\cal R}(r) $, we have
\begin{eqnarray}
\label{eqradpsi}
\dfrac{d^2\chi(r)}{dr^2}+V(r) \chi(r) = 0,
\end{eqnarray}
where
\begin{eqnarray}
\label{poteff}
V(r)=\dfrac{[B'(r)]^2}{4 B^2(r)} - \dfrac{B''(r)}{2B(r)} + \dfrac{\omega^2}{B^2(r)} - 
\dfrac{V_{eff}}{B^2(r)}.
\end{eqnarray}
Now, by expanding the potential  $V(r)$ as a power series in $1/r$ we find
\begin{eqnarray}
V(r) &\approx & \omega^2\left[1 + \dfrac{2 r_{hgup}^{(d-2)} }{r^{(d-2)}} + \dfrac{3 r_{hgup}^{2(d-2)} }{r^{2(d-2)}} + \dfrac{4 r_{hgup}^{3(d-2)} }{r^{3(d-2)}} + \dfrac{5 r_{hgup}^{4(d-2)} }{r^{4(d-2)}} + \dfrac{6 r_{hgup}^{5(d-2)}}{r^{5(d-2)}} + \cdots \right]\nonumber \\ 
&-& \dfrac{l(l+1)}{r^2}\left[ 1 + \dfrac{r_{hgup}^{(d-2)}}{r^{(d-2)}} + \dfrac{r_{hgup}^{2(d-2)}}{r^{2(d-2)}} + \cdots \right] - \dfrac{(d-2)}{r^2}\left[ \dfrac{r_{hgup}^{(d-2)}}{r^{(d-2)}}  + \dfrac{r_{hgup}^{2(d-2)}}{r^{2(d-2)}} + \dfrac{r_{hgup}^{3(d-2)}}{r^{3(d-2)}} + \cdots\right] \nonumber \\
&+& \dfrac{(d-1)(d-2)}{2r^2}\left[ \dfrac{r_{hgup}^{(d-2)}}{r^{d-2}} + \dfrac{r_{hgup}^{2(d-2)}}{2r^{2(d-2)}} + \dfrac{r_{hgup}^{3(d-2)}}{r^{3(d-2)}} + \cdots\right] + \dfrac{(d-2)^2}{4r^2}\left[\dfrac{r_{hgup}^{2(d-2)}}{r^{2(d-2)}} + \dfrac{2r_{hgup}^{3(d-2)}}{r^{3(d-2)}} + \cdots \right] + \cdots .
\end{eqnarray}
We rewrite the radial equation as
\begin{eqnarray}
\dfrac{d^2\chi(r)}{dr^2}+ \left[ \omega^2 + \mathcal{U}(r)\right]\chi(r) = 0,
\end{eqnarray}
where we take the first terms of the expansion of the form
\begin{eqnarray}
\mathcal{U}(r) \approx  \dfrac{12\ell^2}{r^2} - \dfrac{3 r_{hgup}^2}{r^2} + \dfrac{2 \omega^2 r_{hgup}^{(d-2)} }{r^{(d-2)}} + \dfrac{3 \omega^2 r_{hgup}^{2(d-2)} }{r^{2(d-2)}}+ \cdots .
\end{eqnarray}
Here we separate the terms that carry the dimension $d$. We define 
\begin{eqnarray}
\ell^2 = -\dfrac{l(l+1)}{12} + \dfrac{(\omega r_{hgup})^{2}}{4},
\end{eqnarray}
which is the change in the coefficient of $1/r^2$ containing the contributions of $l$, $M$, $\omega$ and the dimension $d$.
It is easy to notice that in the limit $r \rightarrow \infty$ the new potential ${\cal U}_{eff}(r) \rightarrow 0$ satisfying the asymptotic conditions.

Applying the following approximate formula~\cite{Anacleto:2017kmg,Anacleto:2019tdj,Anacleto:2020zhp,Anacleto:2020lel}
\begin{eqnarray}
\label{formapprox}
\delta_l\approx 2(l-\ell) = 2\left(l - \sqrt{-\frac{(l^2+l)}{12} + \dfrac{r_{hgup}^2\omega^2}{4}}\,\right), 
\end{eqnarray}
at the limit $ l\rightarrow 0 $ we have a phase shift $ \delta_{0} \approx - \omega r_{hgup}$.
Thus, we can now determine the differential scattering cross section by considering the following expression~\cite{Yennie:1954zz,Cotaescu:2014jca,dInverno} 
\begin{eqnarray}
\label{espalh}
\dfrac{d\sigma}{d\Omega}=\big|f(\theta) \big|^2=\Big| \frac{1}{2i{\omega}}\sum_{l=0}^{\infty}(2l+1)\left(e^{2i\delta_l} -1 \right)
\frac{P_{l}(\cos\theta)}{1-\cos\theta}\Big|^2.
\end{eqnarray}
Taking the limit for small angles and for $l \rightarrow 0$ we obtain the differential scattering cross section
\begin{eqnarray}
\frac{d\sigma}{d\Omega}\Big |^{\mathrm{l f}}
=\frac{4}{\theta^4}\left[\dfrac{8\Gamma(d/2) M}{(d-1)\pi^{(d-2)/2}} \right]^{2/(d-2)}\left[1 - \dfrac{(d-1)\pi^{(d-2)/2}}{2\Gamma(d/2)} \dfrac{\alpha}{M} + \dfrac{(d-1)^{2}\pi^{(d-2)}}{4\Gamma(d/2)^{2}}\dfrac{\beta}{M^{2}} \right]^{2/(d-2)}.
\label{diffcsect}
\end{eqnarray}

\subsection{Absorption cross section}

\begin{figure}[!htb]
 \centering
 \subfigure[]{\includegraphics[scale=0.35]{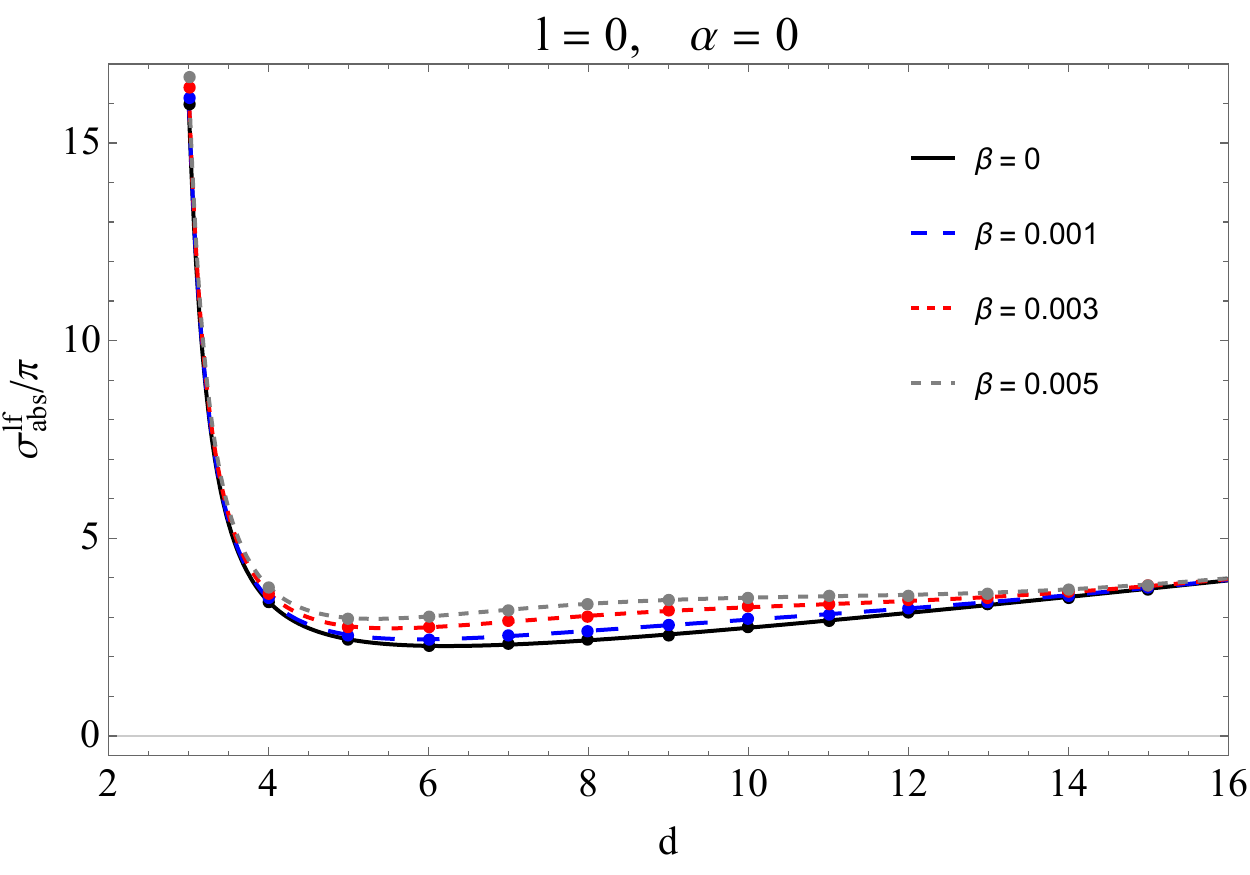}\label{abslowa0}}
 \qquad
 \subfigure[]{\includegraphics[scale=0.35]{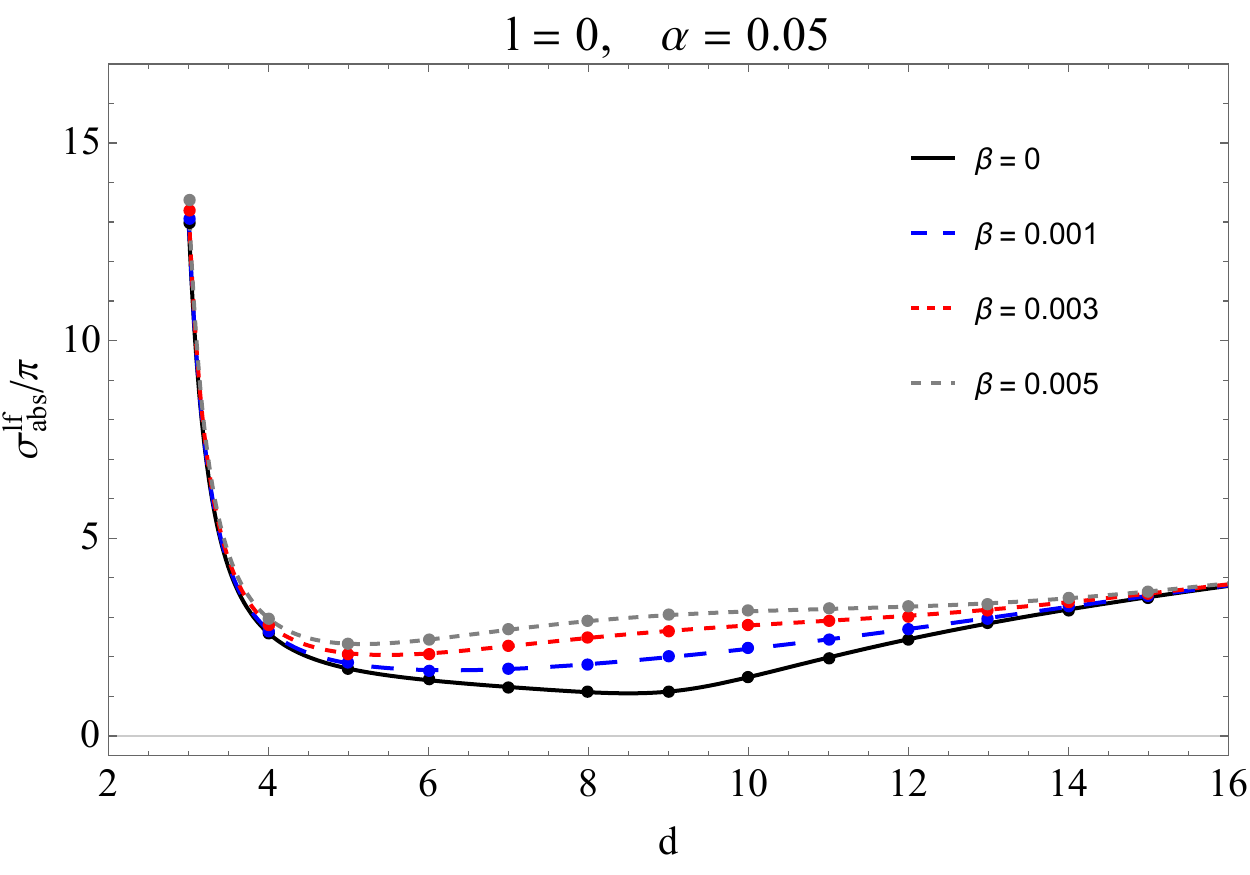}\label{abslowa003}}
 \\
  \caption{\footnotesize{Graphs of the absorption cross section at low frequencies limit  as a function of the number of spatial dimensions. We see that even for very small values of $\beta$, there are changes in the absorption. This is mainly observed from $d = 5$ to $d = 12$. }}
 \label{abslow}
\end{figure}

At this point, we will determine the absorption cross section for the low frequency limit. We can calculate the total absorption cross section with the following relationship
\begin{eqnarray}
\sigma_{abs}
=\frac{\pi}{\omega^2}\sum_{l=0}^{\infty}(2l+1)\Big(1-\big|e^{2i\delta_l}\big|^2\Big)
=\frac{4\pi}{\omega^2}\sum_{l=0}^{\infty}(2l+1)\sin^2(\delta_{l}).
\end{eqnarray}
For the phase shift $\delta_{0}$ found earlier, we can calculate at the low frequency limit $\omega\rightarrow 0$ and $(l=0)$ the following absorption cross section
\begin{eqnarray}
\label{abs1}
\sigma_{abs}^{\mathrm{l f}} = 4\pi r_{hgup}^2 = 4\pi\left[\dfrac{8\Gamma(d/2) M}{(d-1)\pi^{(d-2)/2}} \right]^{2/(d-2)}\left[1 - \dfrac{(d-1)\pi^{(d-2)/2}}{2\Gamma(d/2)} \dfrac{\alpha}{M} + \dfrac{(d-1)^{2}\pi^{(d-2)}}{4\Gamma(d/2)^{2}}\dfrac{\beta}{M^{2}} \right]^{2/(d-2)}.
\end{eqnarray}
Therefore, regardless of the number of spatial dimensions, the absorption cross section is always proportional to the horizon area of the black hole as can be seen in~\cite{Das:1996we}, where a low frequency cross section is calculated for spherically symmetric black holes. Thus, under the aforementioned conditions, it is shown that regardless of the number of dimensions, in a spacetime external to a static and asymptotically flat black hole,  the absorption cross section of the scalar field equals the area of the horizon of the black hole. 
In~\cite{Jung:2004nh}, we also have a result for low frequencies ($l \rightarrow 0$) where the results are equivalent to the absorption cross section \eqref{abs1}.
{In Fig.~\ref{abslow}, we have the behavior for the absorption cross section in the low frequency limit and $l=0$. We can see that for larger dimensions from $d = 5$ to $d = 11$, we have a greater influence of the GUP quantum corrections. In this way, even for $\beta$ of the order of $10^{-3}$, we have a change of the absorption cross section, as we will see in the numerical results.}

\subsection{Null Geodesic}

Some authors,  as for example,  in Refs.~\cite{Singh:2017vfr,Decanini:2011xw} have studied the geodesic for a black hole described by the metric  \eqref{metric2}.
We can find the null geodesics from this metric by taking a Lagrangian of the form 
\begin{eqnarray}
\mathcal{L} \equiv \dfrac{1}{2}g_{\mu\nu}\dot{x}^{\mu}\dot{x}^{\nu}.
\end{eqnarray}
So we have
\begin{equation}
2\mathcal{L} = B(r)\dot{t}^{2} - \dfrac{\dot{r}^{2}}{B(r)} - \sum_{i=1}^{d-2}r^{2}\prod_{n=1}^{i-1} \sin^2\theta_{n} \dot{\theta}_{d-2}^{2} - r^2 \prod_{i=1}^{d-2} \sin^{2}\theta_{i} \dot{\phi}_{d-2}^{2}, 
\label{elidot}
\end{equation} 
where ``." is the derivative with respect to the affine parameter.

Here, we are interested in the trajectory of a light ray in the described metric which is spherically symmetric. 
So, if we analyze in a plane, any ray of light that starts with a certain angle $\theta$ must remain with the same angle.
 We will then consider an equatorial plane fixing the angle $\theta_{i}$ at $\pi/2$.

\begin{figure}[!htb]
 \centering
 \subfigure[]{\includegraphics[scale=0.35]{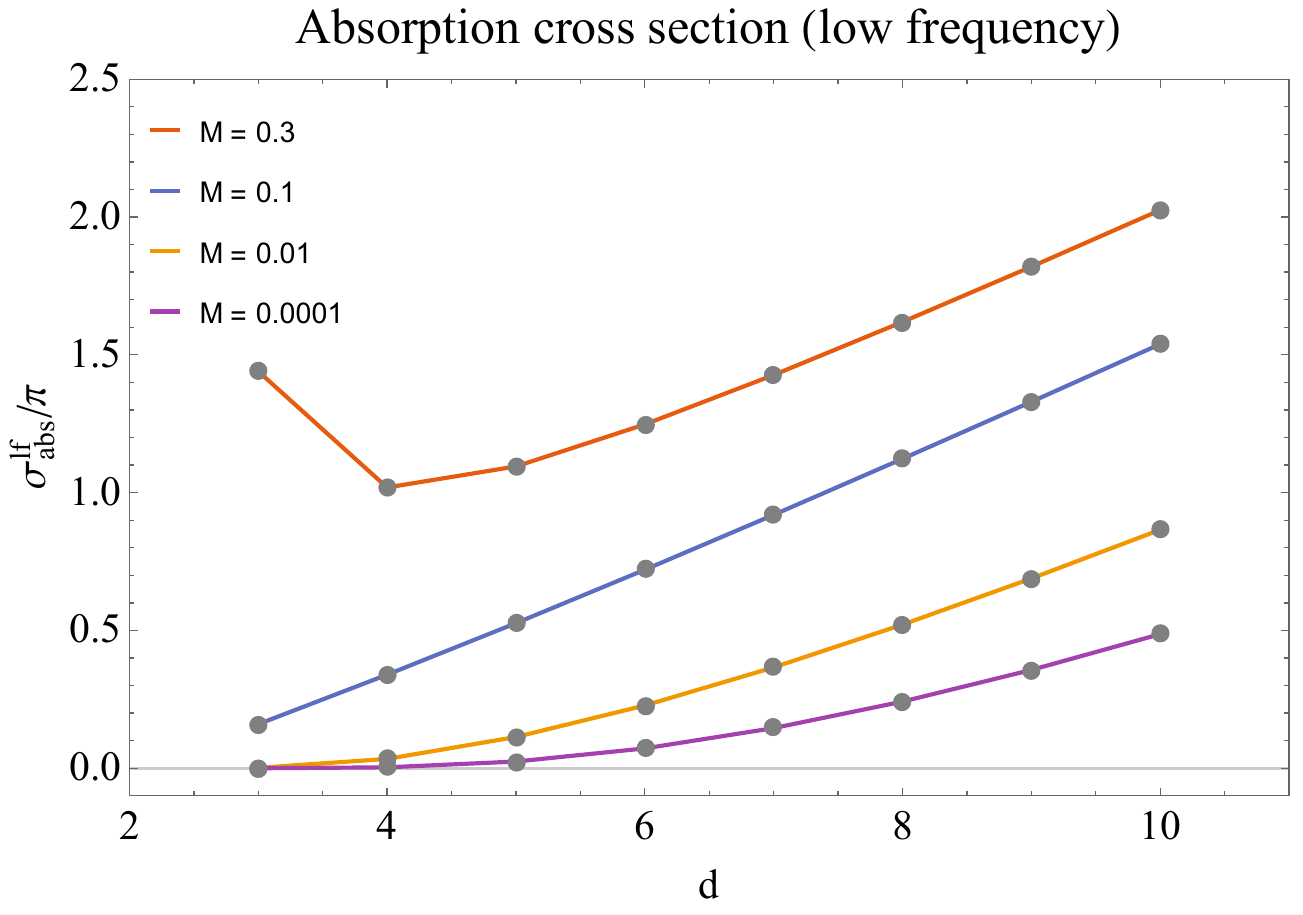}\label{abslowf}}
 \qquad
 \subfigure[]{\includegraphics[scale=0.35]{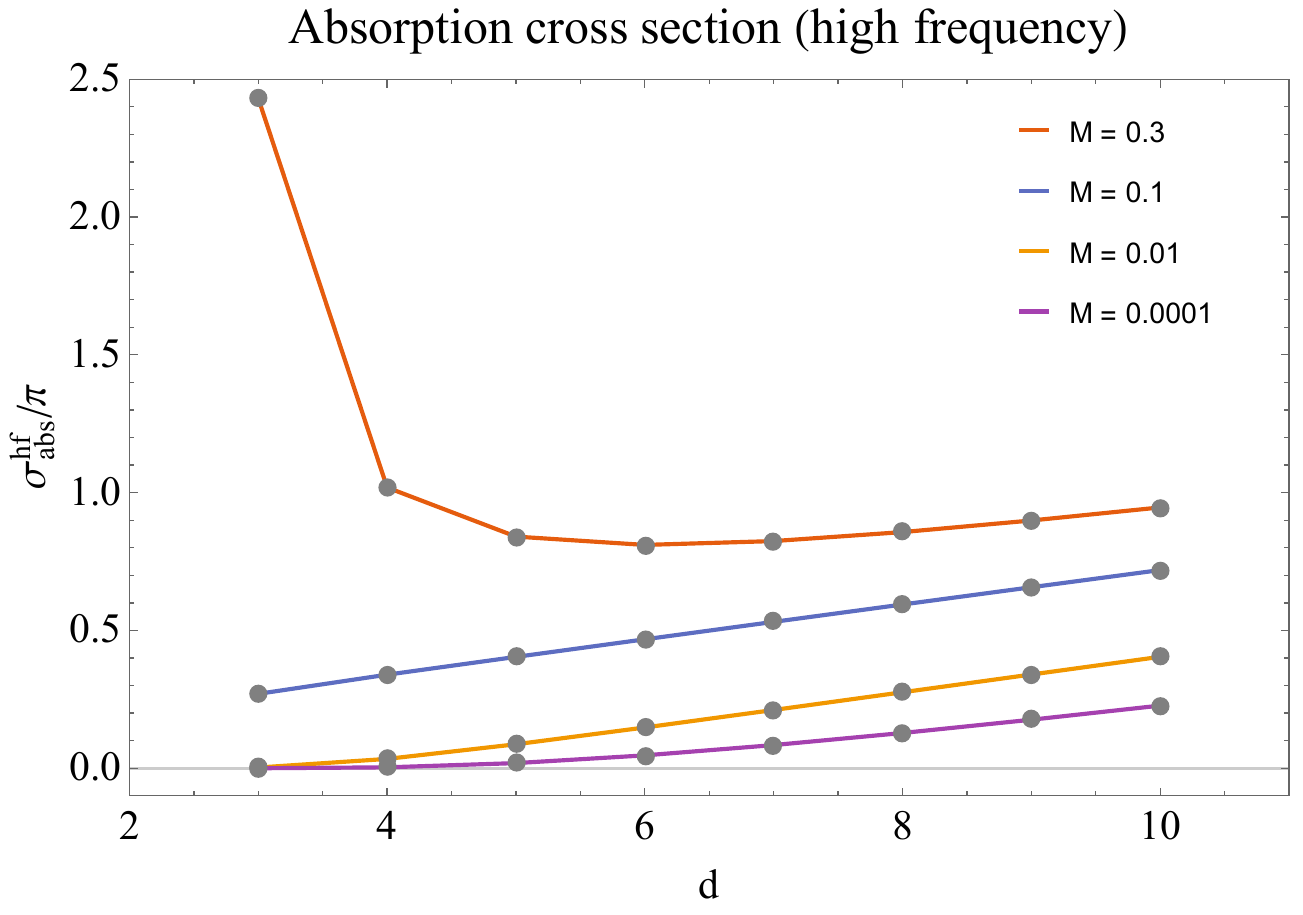}\label{abshighf}}
 \\
  \caption{\footnotesize{We can see that even for a limit of small masses $M \rightarrow 0$ we have the presence of absorption in high dimensions. }}
 \label{abslowhi}
\end{figure}

\begin{figure}[!htb]
 \centering
\includegraphics[scale=0.45]{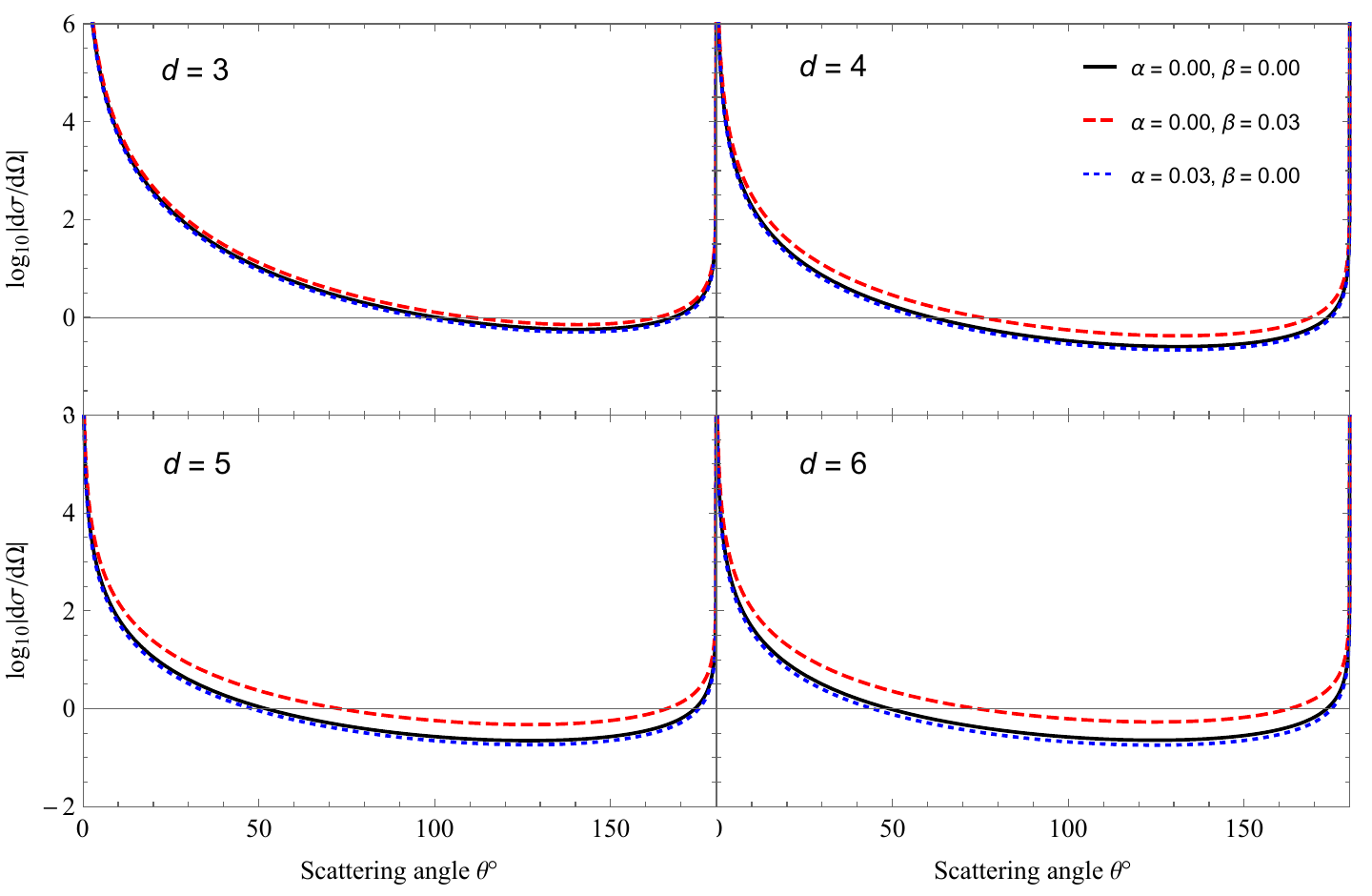}\\
  \caption{\footnotesize{Scattering cross section in the classical limit for spatial dimensions $4$, $5$ and $6$. The black curve is the conventional case $\alpha = 0$ and $\beta$ while the dashed red curve is the contribution of the quadratic part only and the dotted blue one is the linear part only.}}
 \label{ClassPlot}
\end{figure}

Furthermore, under these conditions \eqref{elidot} is independent of $t$ and $\phi$. So two equations are enough to describe the motion of a ray of light. We can build a system with these equations that give rise to two constants of geodetic motion $E$ and $L$, which correspond to energy and angular momentum respectively:
\begin{equation}
E = B(r)\dot{t}, \qquad \quad L = r^{2}\dot{\phi}.
\label{EL}
\end{equation}
For null geodesics we have $g_{\mu\nu}\dot{x}^{\mu}\dot{x}^{\nu} = 0$, and using the equations \eqref{EL}, we can write the equation of ``energy"
\begin{equation}
\dot{r}^{2}  + B(r)\dfrac{L^{2}}{r^{2}} = E^{2}.
\label{eqEner}
\end{equation}
So far the results are identical to those obtained for the Schwarzschild case~\cite{dInverno}. 
However, the difference is in the metric function $B(r)$, 
where there are informations about extra dimensions.
Hence, introducing a new variable $u = 1/r$ we can write the orbital equation as follows
\begin{eqnarray}
&\dfrac{du}{d\phi} =\sqrt{\dfrac{1}{b^2} - u^{2} + r_{hgup}^{d-2} u^{d}} \label{eqD1},\\
&\dfrac{d^{2}u}{d\phi^{2}} + u = d\dfrac{r_{hgup}^{d-2}}{2}  u^{d-1},
\label{eqD2}
\end{eqnarray}
where $b = L/E$ is the impact parameter defined as the perpendicular distance (measured at infinity) between the geodesic and a parallel line passing through the origin. The second order differential equation is the relativistic Binet's equation. Hence, by analyzing the critical case where 
\begin{eqnarray}
\dfrac{du}{d\phi} = 0, \qquad \dfrac{d^{2}u}{d\phi^2} = 0,
\end{eqnarray}
we find the critical radius 
\begin{eqnarray}
r_c^{d-2}=\frac{d}{2}r_{hgup}^{d-2}=d M_{gup}^{d-2}
=\frac{4d\,\Gamma\left({d}/{2}\right)M}{(d-1)\pi^{(d-2)/2}}\left[1 - \dfrac{(d-1)\pi^{(d-2)/2}}{2\Gamma(d/2)} \dfrac{\alpha}{M} + \dfrac{(d-1)^{2}\pi^{(d-2)}}{4\Gamma(d/2)^{2}}\dfrac{\beta}{M^{2}} \right].
\end{eqnarray}
Note that for $d=3$, we have $r_c=3M\left[1- 2\alpha/M + 4\beta/M^2  \right]$ which is the result obtained in~\cite{Anacleto:2021qoe}.
The critical radius describes an unstable orbit where the photo beam orbits the black hole. Thus,  we can also calculate the critical impact parameter in terms of the critical radius
\begin{equation}
b_{c} = \dfrac{r_{c}^{(d+2)/2}}{\sqrt{r_{c}^d - r_{hgup}^{d-2}r_{c}^{2}}}.
\label{bcrit}
\end{equation}
For $d=3$ the result obtained in~\cite{Anacleto:2021qoe} is recovered, namely $b_c=3\sqrt{3}M\left[1- 2\alpha/M + 4\beta/M^2  \right]$.
In Fig.~\ref{abslowhi}, we can see the behavior of the absorption cross section for low and high frequency as a function of dimension for $\alpha = \beta = 0$. Notice that even for small $M$, in higher dimensions the absorption is not zero. Also, we can see a variation when comparing the case \ref{abslowf} with the isotropic case that, even in $l = 0$, the higher dimensional aspects guarantees non zero absorption cross section. This also occurs, but more smoothly, for high energies~\ref{abshighf}. 
\\
By using the equations \eqref{eqD1} and \eqref{eqD2}, we derive an analytical approximation for the differential scattering cross section for small angles.
The approximate deflection angle as a function of the impact parameter is
\begin{equation}
\Theta(b) \approx d\cdot F\left(\dfrac{1}{2}, -\dfrac{(d - 1)}{2}; \dfrac{3}{2}; 1\right) \left( \frac{r_{hgup}}{b}\right)^{d-2},
\label{angdef}
\end{equation}
where we again have the result in terms of a hypergeometric function $F$.
Therefore, we have the following results for the approximate deflection angle
\begin{eqnarray}
\Theta(b) \approx  \left\{ \dfrac{4M_{gup}}{b}, \dfrac{2M_{gup}^2}{b^2}, \dfrac{4M_{gup}^3}{\pi b^3}, \dfrac{3M_{gup}^4}{\pi b^4},\dfrac{8M_{gup}^5}{\pi b^5}\right\},
\label{angdef2}
\end{eqnarray}
{for the respective values of $d = \lbrace 3, 4, 5, 6, 7 \rbrace $.
Note that, as expected, for $d=3$, $\alpha = 0$, and $\beta = 0$, we return to the conventional Schwarzschild case, and we recover the Einstein deflection angle. Next, we will check the other results for the deflection angle by using the limit of large $l$. Then, the classical differential scattering cross section is given by}
\begin{equation}
\dfrac{d\sigma}{d\Omega} \Big|_{cl} = \dfrac{b}{\sin \theta} \Big|\dfrac{db}{d\theta} \Big|.
\label{scatcl}
\end{equation}
By applying  the equations \eqref{angdef} and \eqref{scatcl} the differential scattering cross section for small angles is given by
\begin{footnotesize}
\begin{equation}
\dfrac{d\sigma}{d\Omega} \Big|_{cl} \approx \dfrac{\left[d\cdot F\left(\dfrac{1}{2}, \dfrac{(1-d)}{2}; \dfrac{3}{2}; 1\right)\right]^{2/(d-2)}}{\pi(d - 2)\theta^{2(d-1)/(d-2)}} \left[\dfrac{8 M \Gamma(d/2)}{(d-1)} \right]^{2/(d-2)}
\left[1 - \dfrac{(d-1)\pi^{(d-2)/2}}{2\Gamma(d/2)} \dfrac{\alpha}{M} + \dfrac{(d-1)^{2}\pi^{(d-2)}}{4\Gamma(d/2)^{2}}\dfrac{\beta}{M^{2}}  \right]^{2/(d-2)}.
\label{scattClass} 
\end{equation}
\end{footnotesize}
In Eq.~\eqref{scattClass}, we have that the differential scattering cross section decreases for higher dimensions. On the other hand, the effect of the correction coming from the GUP can be seen in Fig.~\ref{ClassPlot}. The contribution of the quadratic part $\beta$ (dashed red line) is more relevant shifting the scattering related curves upwards in higher dimensions.

{The classic formula \eqref{scattClass} does not consider high angles. 
In this way, it is interesting to present an approximate method that works well for high scattering angles, specifically close to $\pi$. This method is called the glory semi-classical approximation~\cite{matzner1985glory}. The advantage of this semi-classical approximation is to obtain an analytical formula by giving a physical view of the width of the interference fringes in the differential scattering cross section and the intensity of the scattered flux for $\theta \sim \pi$. Although a glorified semi-classical approximation is valid for high frequencies $\omega \gg 1$, it still agrees with the numerical results for intermediate frequencies $\omega \sim 1$~\cite{crispino2009scattering}.
The semi-classical formula for glory scattering for a spherically symmetric black hole is given by \cite{matzner1985glory}}
\begin{eqnarray}
\dfrac{d\sigma}{d \Omega} = 2\pi \omega b_{g}^{2} \Big | \dfrac{d b}{d \theta}  \Big |_{\theta = \pi} J_{2s}^{2}\left(\omega b_{g}\sin\theta \right),
\label{scatglory}
\end{eqnarray}
{where $b_{g}$ is the impact parameter for the backscattered rays ($\theta = \pi$), $J_{2s}(x)$ is the Bessel function of the first kind, and $s$ is the spin of the wave. We analyze here only the scalar case where $s = 0$. There are several values of $b_{g}$ that correspond to multiple values of the deflection angle ($\chi = \pi + 2n\pi$ with $n = 0, 1, 2 \ldots$),  the number of times the light beams revolve around the black hole, but the biggest contribution to the differential scattering cross section comes from $n=0$~\cite{macedo2015scattering}. So according to the semi-classical equation for glory scattering \eqref{scatglory}, we have to get $b_{g}$ and $|db/d\theta|_{\theta=\pi}$. The value of $b_{g}$ can be obtained numerically by solving the orbital equation \eqref{eqD1}, and later, we obtain $|db/d\theta|_{\theta=\pi}$. In Table \ref{tab1}, we can see some results obtained with the increase of the dimension.  The glory parameter $b_{g}$ is very close to the critical impact parameter \eqref{bcrit}.}
\begin{table}[!ht]
\begin{center}
\caption{\footnotesize{Parameters of the glory approximation obtained numerically and critical impact parameter of the dimensions $4$ and $6$. }} 
\label{tab1}
\begin{tabular}{|c||c|c|c|c|c|c|}
\hline
\multicolumn{1} {|c||}{$\alpha = 0$} & \multicolumn{3} {|c|}{ $d = 4$ } & \multicolumn{3} {|c|}{ $d = 6$ }   \\
\hline
 $\beta$     & $b_{c}$ &  $b_{g}$ & $|db/d\theta|_{\theta=\pi}$  & $b_{c}$ &  $b_{g}$ & $|db/d\theta|_{\theta=\pi}$\\
 \hline
 0     & 1.84264  & 1.8509 &  0.11894$\times 10^{-1}$ & 1.21629 & 1.21675  & 0.9206$\times 10^{-3}$     \\
 0.03  & 2.37850  & 2.3892 &  0.15853$\times 10^{-1}$ & 1.86821 & 1.86891  & 1.4130$\times 10^{-3}$      \\
  \hline
 \end{tabular}
\end{center}
\end{table}
{According to the glory effect formula, the distance between the peaks of the differential scattering cross section varies inversely with the parameter $b_{g}$. So for each dimension, the parameter $\beta$ of the GUP correction influences the scattering cross section decreasing the distance between the peaks, as we can see in Table \ref{tab1} and in Figs. \ref{scatt} and \ref{glory}.}

Now, to numerically calculate the absorption cross section, it is enough to solve the radial equation \eqref{eqrad}, by considering the boundary conditions and then relate them to the transmission and reflection coefficients. For the differential scattering cross section it is a little more complicated due to the need of adding large contributions of partial waves. In this case the contribution of the Legendre polynomial in the phase amplitude formula can generate errors in the final result of the scattering cross section.
Thus, to solve this problem we use a convergence method~\cite{Yennie:1954zz} based on the recurrence relations of Legendre polynomials.
However, from the $d\geq6$ dimensions we can obtain precise results without using the convergence method, by adding to the numerical results a semi-classical contribution as done in~\cite{Dolan:2009zza}. Thus, to obtain the results of the scattering cross section and absorption, we will use an approximation for the phase shift in the very large $l$ regime. We start by rewriting the equation \eqref{metric2} as follows
\begin{eqnarray}
\frac{d^2\chi(r)}{dr^2}+\left[\omega^2 - \dfrac{l(l+1)}{r^2} - U(r)\right] \chi(r) = 0,
\end{eqnarray}
where the effective potential is now given by
\begin{eqnarray}
U(r) &=& \left[-2\omega^2 + \dfrac{2l(l+1) - (d-3)(d-2) }{2r^{2}} \right]\dfrac{r_{hgup}^{d-2}}{r^{d-2}} + \left[-3\omega^2 + \dfrac{2l(l+1) - (d-3)(d-2) }{2r^{2}} \right]\dfrac{r_{hgup}^{2(d-2)}}{r^{2(d-2)}} + \cdots.
 \label{pot1}
\end{eqnarray}
{Here we truncate the series at $1/r^{d-2}$, where $d\geq 4 $.} Next, we apply the approximate Born formula~\cite{Morse:1954},
\begin{eqnarray}
\delta_{l} \approx - \omega \int_{0}^{\infty}r^2 J_{l}^{2}(\omega r)U(r)dr,
\end{eqnarray}
where $J_{l}(\omega r)$ is the Bessel function of the first kind. { We arrive at the following expression for the phase shift}
\begin{eqnarray}
\delta_{l} \approx -\sqrt{\pi}\left(\omega r_{hgup}\right)^{d-2} \left[l(l+1)(1-d) + (d-3)(d-2) \right]\dfrac{\Gamma\left(\dfrac{d-3}{2}\right)\Gamma\left(l-\dfrac{d-3}{2}\right)}{8\Gamma\left(\dfrac{d}{2}\right)\Gamma\left(\dfrac{1+d}{2}+l\right)}.
\end{eqnarray}
Next, we calculate the phase shift for some values of $d$ using the first terms of each respective potential $U(r)$ and applying the following identities:
\begin{eqnarray}
\int_{0}^{\infty}J_{l}^{2}(x)x^{-1}dx &=& \dfrac{1}{2l(l + 1)}, \nonumber \\
\int_{0}^{\infty}J_{l}^{2}(x)x^{-2}dx &=& \dfrac{\pi}{(2l - 1)(2l + 1)(2l + 3)}, \nonumber \\
\int_{0}^{\infty}J_{l}^{2}(x)x^{-3}dx &=& \dfrac{1}{3l(l - 1)(l + 1)(l + 2)}, \nonumber \\
\int_{0}^{\infty}J_{l}^{2}(x)x^{-4}dx &=& \dfrac{\pi}{(2l - 3)(2l - 1)(2l + 1)(2l + 3)(2l + 5)}.
\end{eqnarray}
In this way, we obtain the following results for the phase shifts expressed in terms of $l + 1/2$ for the respective values $d={4, 5, 6, 7}$
\begin{eqnarray}
\delta^{(d=4)}_{l} &\approx& \dfrac{3\pi\left(\omega r^{(d=4)}_{h}\right)^2}{8(l + 1/2)} = \dfrac{M_{gup}^2 \omega^{2}}{(l + 1/2)}, \qquad e^{2i\delta^{(d=4)}_{l}} \sim  1 + \dfrac{2i M_{gup}^2 \omega^{2} }{(l + 1/2)},
\label{deltSC4}\\
\delta^{(d=5)}_{l} &\approx& \dfrac{2\left(\omega r^{(d=5)}_{h}\right)^3}{3(l + 1/2)^2} = \dfrac{M_{gup}^3 \omega^{3}}{\pi(l + 1/2)^{2}}, \qquad e^{2i\delta^{(d=5)}_{l}} \sim 1 + \dfrac{2i M_{gup}^3 \omega^{3} }{\pi(l + 1/2)^{2}},\\ 
\label{deltSC5}
\delta^{(d=6)}_{l} &\approx& \dfrac{5\pi\left(\omega r^{(d=6)}_{h}\right)^4}{32(l + 1/2)^3} = \dfrac{M_{gup}^4 \omega^{4}}{2\pi(l + 1/2)^{3}}, \qquad e^{2i\delta^{(d=6)}_{l}} \sim 1 +  \dfrac{i M_{gup}^4 \omega^{4} }{\pi(l + 1/2)^{3}},\\
\label{deltSC6}
\delta^{(d=7)}_{l} &\approx& \dfrac{2\left(\omega r^{(d=7)}_{h}\right)^5}{5(l + 1/2)^4} = \dfrac{M_{gup}^5 \omega^{5}}{\pi^{2}(l + 1/2)^{4}}, \qquad e^{2i\delta^{(d=7)}_{l}} \sim 1 +  \dfrac{2i M_{gup}^5 \omega^{5} }{\pi^{2}(l + 1/2)^{4}}. 
\label{deltSC7}
\end{eqnarray}
{A semi-classical description of scattering can be found in~\cite{Ford:2000uye}. The impact parameter $b$ is associated with each partial wave as $b = (l + 1/2)/\omega$. Thus, a deflection function has been found that can be related to the deflection angle of a weak field. Notice the case $d=4$ dimension as an example}
\begin{eqnarray}
\Theta = -\dfrac{d}{dl}\left[ Re(2\delta_{l})\right]  = \dfrac{3\pi(\omega r_{hgup})^2}{4(l + 1/2)^2}\Big|_{d=4}.
\end{eqnarray}
{In this way, we have the following angles of deflection}
\begin{eqnarray}
\Theta^{(d=4)} = \dfrac{2 M_{gup}^2}{b^2}, \qquad \Theta^{(d=5)} = \dfrac{4 M_{gup}^3}{\pi b^3} , \qquad \Theta^{(d=6)} = \dfrac{3 M_{gup}^4}{\pi b^4}, \qquad \Theta^{(d=7)} = \dfrac{8 M_{gup}^5}{\pi^{2}b^5}.
\label{Theta}
\end{eqnarray}
{Note that this result is equivalent to the deflection angle \eqref{angdef2}.} 
{Next, we will make a numerical analysis of the results obtained.}

\section{Numerical Analyses }
\label{na}

\begin{figure}[!htb]
 \centering
 \subfigure[]{\includegraphics[scale=0.35]{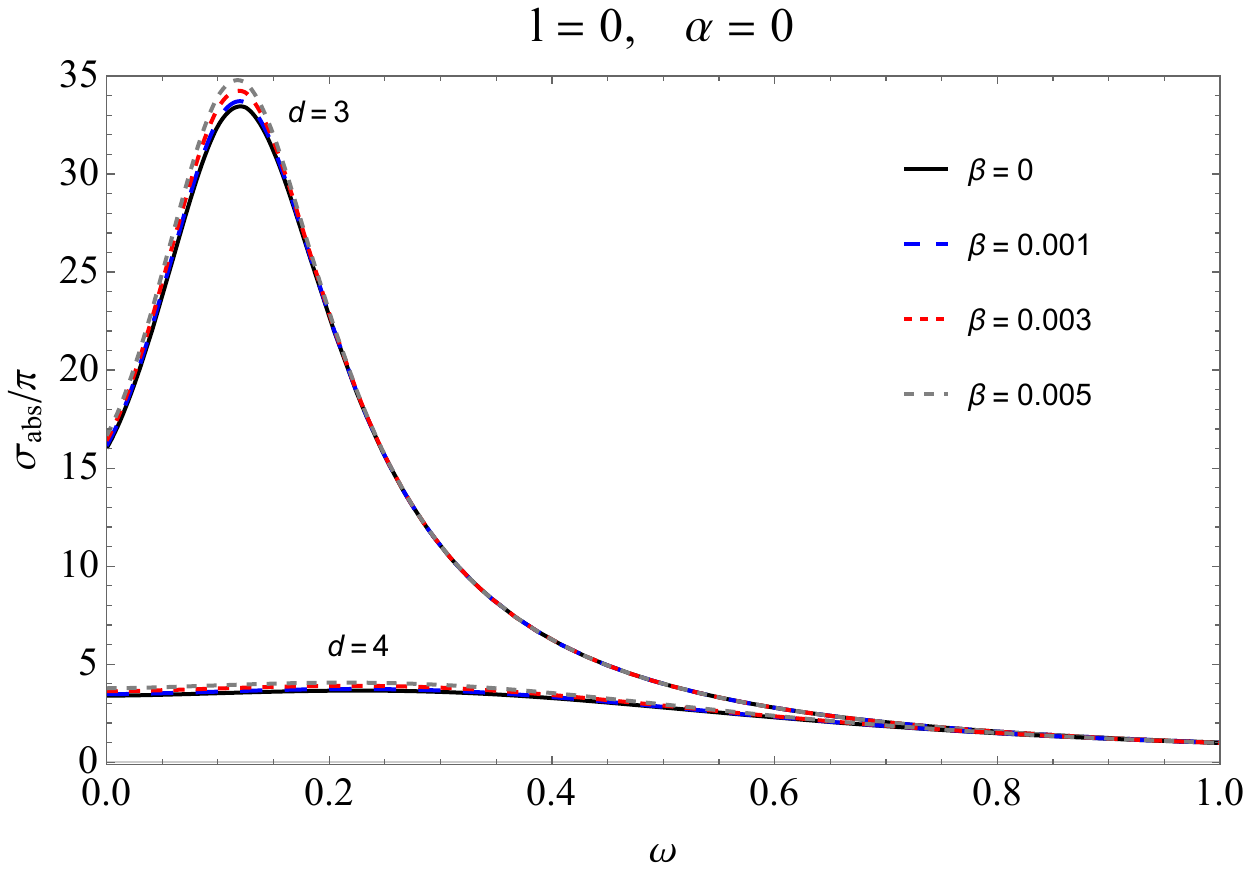}\label{absl0d3d4}}
 \qquad
 \subfigure[]{\includegraphics[scale=0.35]{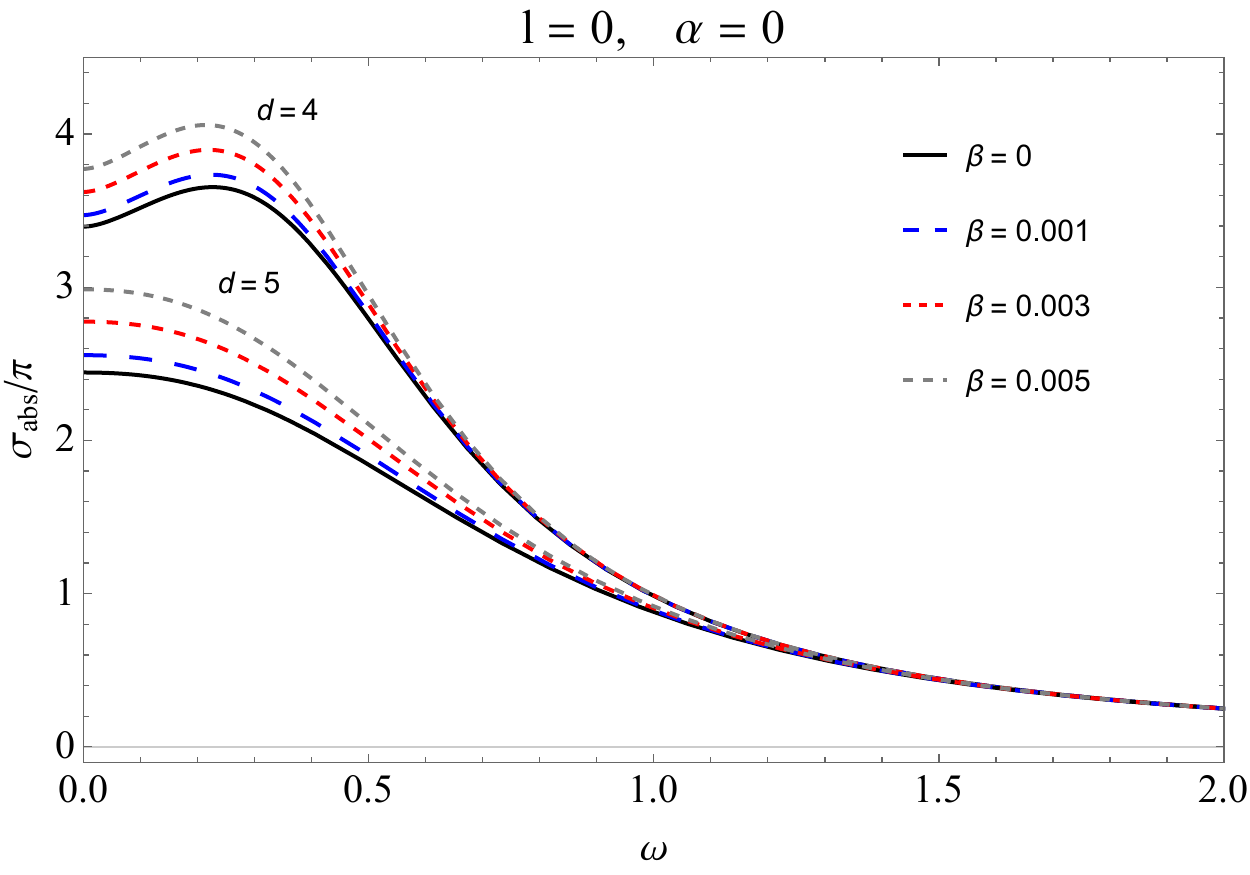}\label{absl0d4d5}}
 \\
  \caption{\footnotesize{Absorption cross section for $l=0$ and $M=1$, assuming only the quadratic part of the GUP, that is, $\alpha = 0$ and $\beta$ varying. For the dimensions $d=3$ and $d=4$ in (a), we can see the GUP influence. In (b) we have $d=4$ and $5$.}}
 \label{absl0}
\end{figure}

\begin{figure}[!htb]
 \centering
 \subfigure[]{\includegraphics[scale=0.35]{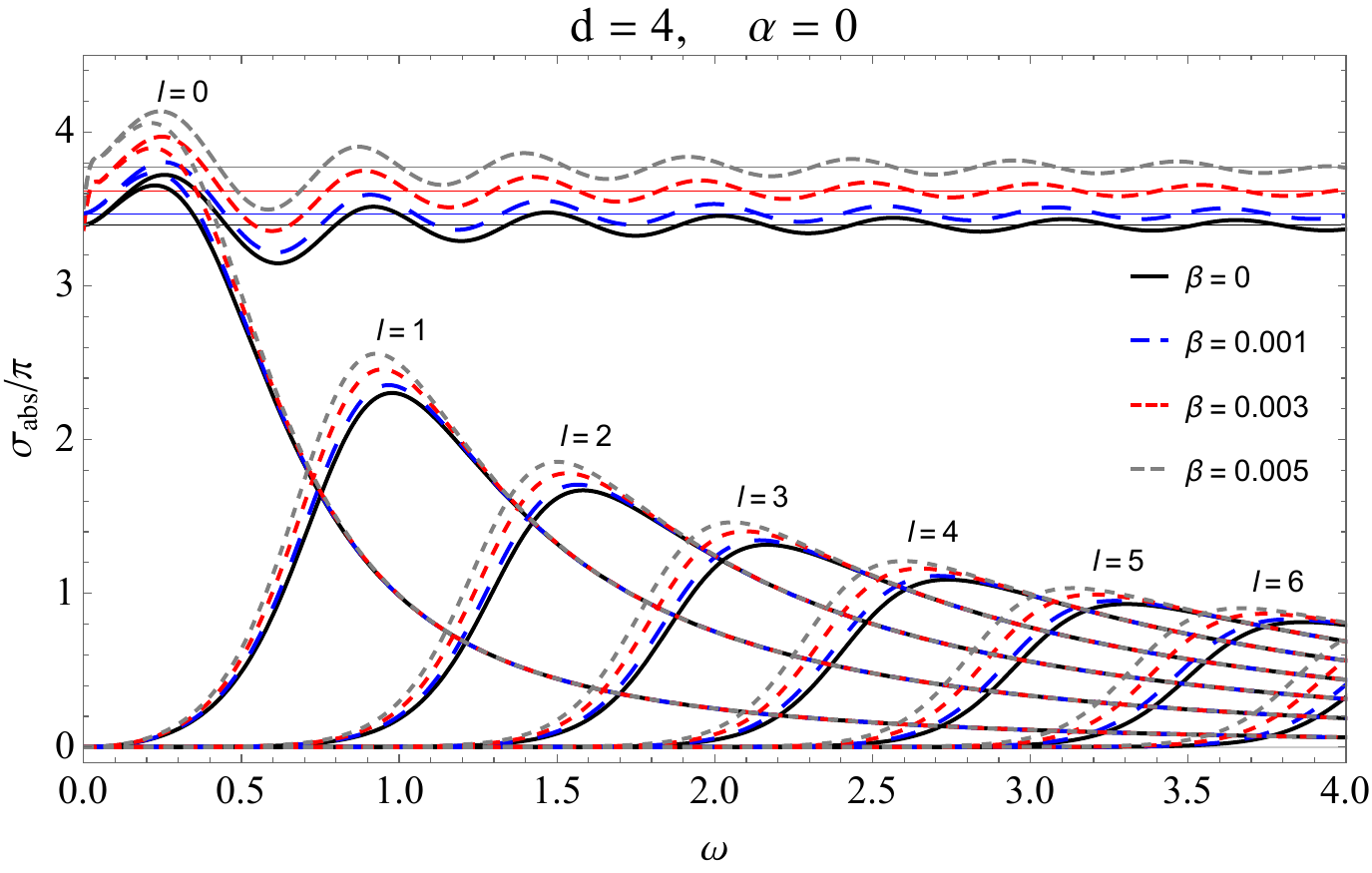}\label{abstotald4}}
 \qquad
 \subfigure[]{\includegraphics[scale=0.35]{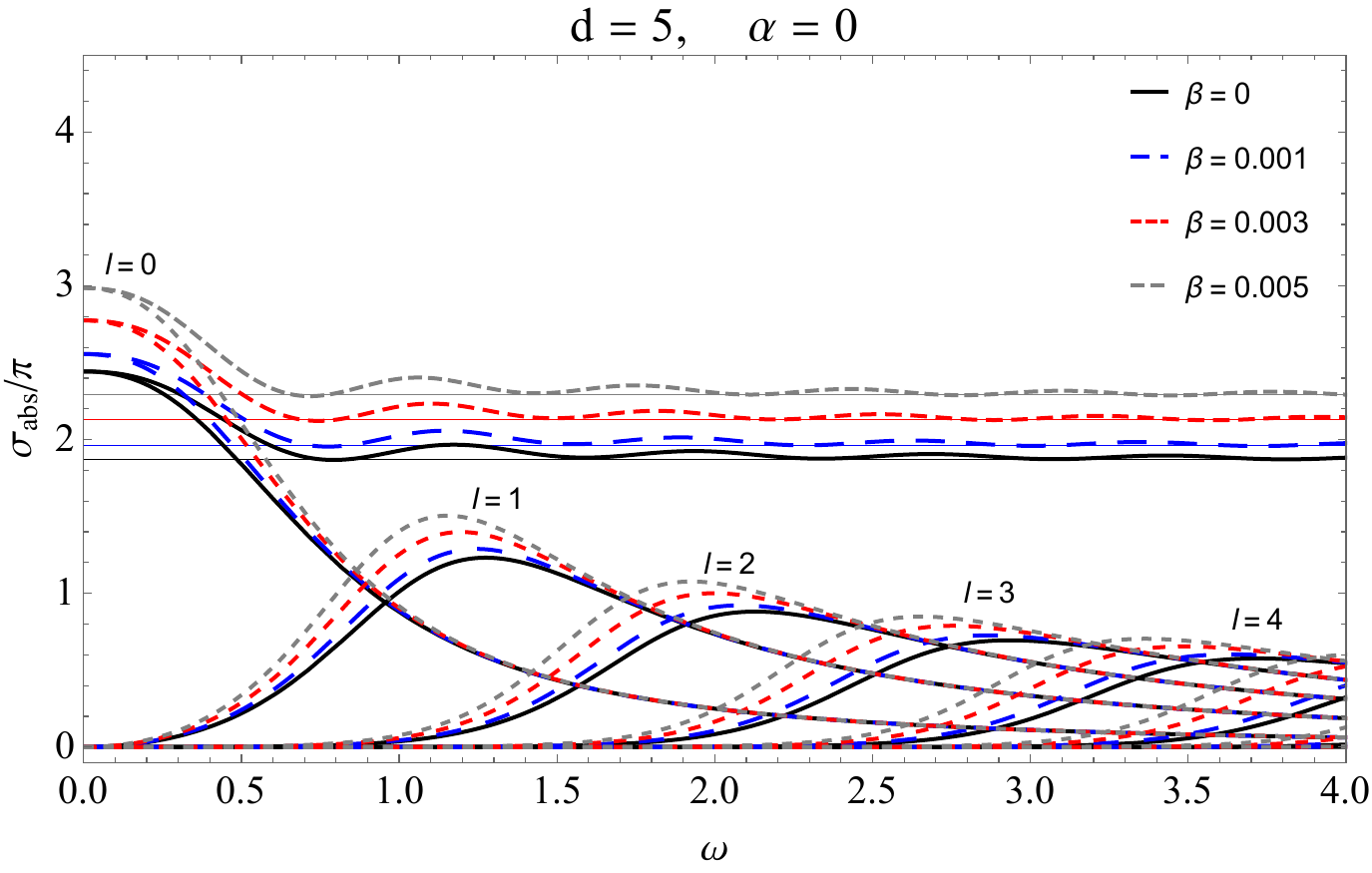}\label{abstotald5}}
 \qquad
 \subfigure[]{\includegraphics[scale=0.35]{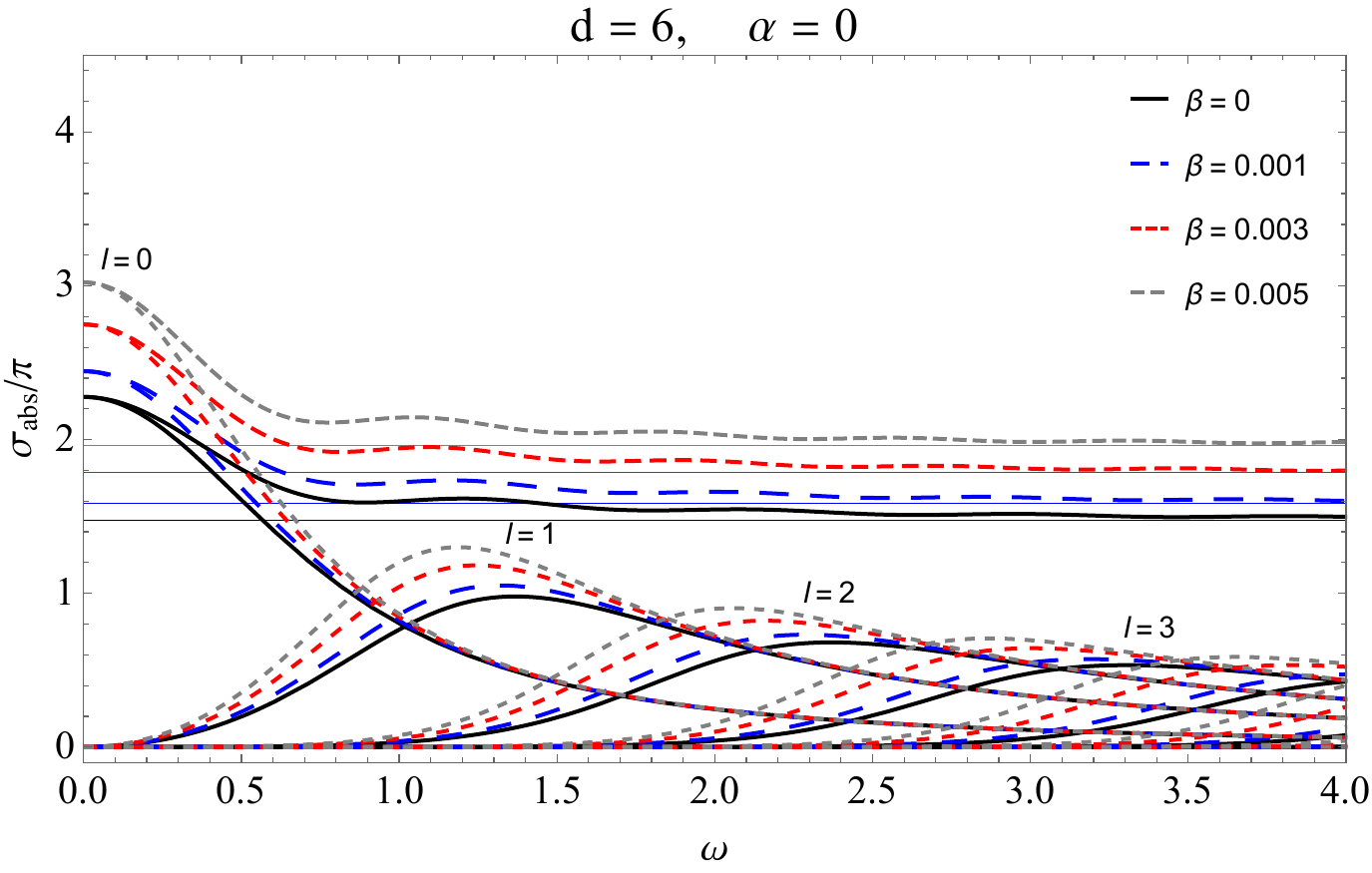}\label{abstotald6}}
  \qquad
 \subfigure[]{\includegraphics[scale=0.35]{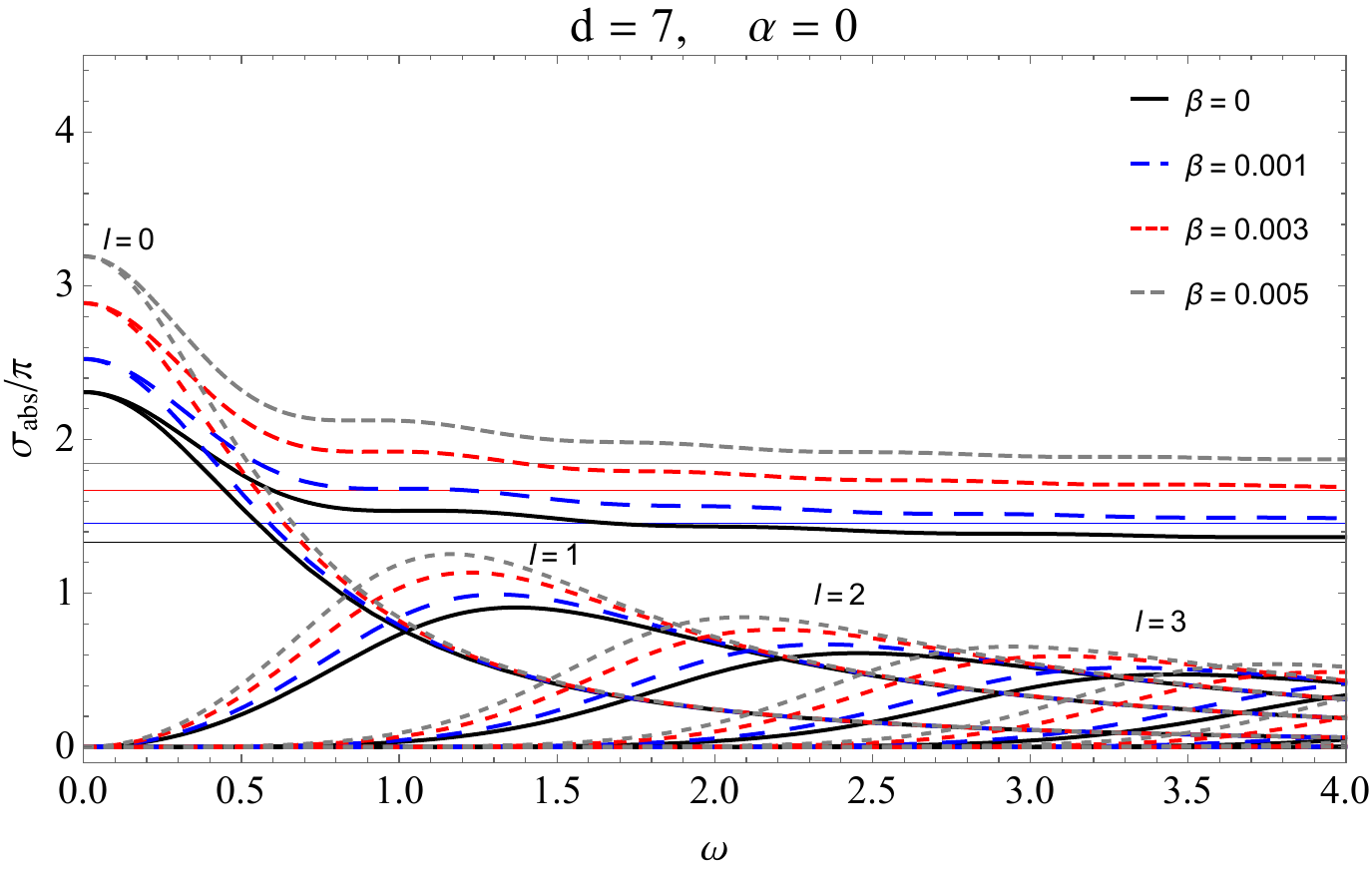}\label{abstotald7}}
 \\
  \caption{\footnotesize{Partial and total absorption cross section from $d=4$ to $d=7$, by assuming $M = 1$. In each figure, we have the sum of the partial absorption that converge to the value of the absorption cross section at high frequencies, which can be confirmed with the geodesic result that is represented in each graph by the straight lines.}}
 \label{abstotald4d6}
\end{figure}

{The table below has some numerical results compared with the analytical equation \eqref{abs1}, for the absorption cross section, in the dimensions $d = 3, 4, 5, 6$, and $7$, fixing the values of $M = 1$ and $l = 0$. For $d = 3$, we have the already known result for Schwarzschild $\sigma_{abs} = 16\pi$ with $\alpha = 0$ and $\beta = 0$. For the other dimensions, the absorption cross section at low frequencies has a considerable reduction in its magnitude. In Table \ref{tab2}, we verify the behavior of the absorption cross section for two values of $\alpha$, i.e., zero and $0.03$; and varying the parameter $\beta$ on a small scale, here as in Fig.~\ref{absl0}, we have the behavior of the absorption cross section in relation to the parameters of GUP corrections, and the numerics matches the analytical approximation.}
\begin{table}[!ht]
\begin{center}
 \begin{footnotesize}
\caption{\footnotesize{Values for the absorption cross section for $\omega \rightarrow 0$ and $l = 0$. }} 
\label{tab2}
\begin{tabular}{|c||c|c|c|c|c|c|c|c|c|c|}
\hline
\multicolumn{1} {|c||}{$\alpha = 0$}&\multicolumn{2} {|c|}{ $d = 3$ } & \multicolumn{2} {|c|}{ $d = 4$  } & \multicolumn{2} {|c}{ $d = 5$  } & \multicolumn{2} {|c}{ $d = 6$  } & \multicolumn{2} {|c|}{ $d = 7$  } \\
\hline
 $\beta$     & eq \eqref{abs1} &  numerical & eq \eqref{abs1}  & numerical & eq \eqref{abs1}  & numerical & eq \eqref{abs1}  & numerical & eq \eqref{abs1}  & numerical\\
 \hline
 0.000  & 16.0000  & 16.0001  & 3.39531  & 3.39530  & 2.44355 &  2.44350  & 2.27764 & 2.27766 & 2.30949 & 2.30946 \\
 0.001  & 16.1283  & 16.1285  & 3.47070  & 3.47070  & 2.55658 &  2.55653  & 2.44483 & 2.44486 & 2.52460 & 2.52456 \\
 0.003  & 16.3863  & 16.3864  & 3.62150  & 3.62150  & 2.77551 &  2.77547  & 2.74888 & 2.74891 & 2.88764 & 2.88759 \\
 0.005  & 16.6464  & 16.6466  & 3.77230  & 3.77229  & 2.98612 &  2.98607  & 3.02250 & 2.02253 & 3.19260 & 3.19255 \\
 \hline
\multicolumn{1} {|c||}{$ \alpha = 0.03$}&\multicolumn{2} {|c|}{ $d = 3$ } & \multicolumn{2} {|c|}{ $d = 4$  } & \multicolumn{2} {|c}{ $d = 5$  } & \multicolumn{2} {|c}{ $d = 6$  } & \multicolumn{2} {|c|}{ $d = 7$  } \\
\hline
 $\beta$     & eq \eqref{abs1} &  numerical & eq \eqref{abs1}  & numerical & eq \eqref{abs1}  & numerical & eq \eqref{abs1}  & numerical & eq \eqref{abs1}  & numerical\\
\hline
 0.000  & 14.1376  & 14.1377  & 2.91531  & 2.91530  & 2.01472 & 2.01468  & 1.80766 &  1.80769 & 1.78647 & 2.78645 \\
 0.001  & 14.2582  & 14.2583  & 2.99070  & 3.99070  & 2.13875 & 2.13870  & 2.01425 &  2.01428 & 2.08630 & 2.08626 \\
 0.003  & 14.5009  & 14.5009  & 3.14150  & 3.14150  & 2.37666 & 2.37662  & 2.37410 &  2.37413 & 2.54468 & 2.54464 \\
 0.005  & 14.7456  & 14.7456  & 3.29230  & 3.29229  & 2.60321 & 2.60316  & 2.68617 &  2.68619 & 2.90408 & 2.90403 \\
 \hline
 \end{tabular}
 \end{footnotesize}
\end{center}
\end{table}
\\
{In Fig.~\ref{absl0}, we have the absorption cross section for $l = 0$. In Fig \ref{absl0d3d4}, we can see the scaling relationship between the absorption for $d = 3$ and $d = 4$ and the influence of the quadratic part of the GUP. Even on a small scale, the behavior of the other dimensions, Figure \ref{absl0d4d5} presents the same for $d=3$, i.e., the absorption cross section does not go to zero, for the isotropic case $\omega \rightarrow 0$ in $l=0$ which does not happen with the other modes $l>0$, as can be seen in figure \ref{abstotald4d6}.} {The total absorption cross section is shown in Fig.~\ref{abstotald4d6}, where we have, the effect of the correction parameter $\beta$ for higher dimensions. We can see that, even for a small value of $\beta$, we have an increase in the absorption cross section for each dimension. The result of the total absorption cross section is confirmed by the geometric-optics value $\sigma_{abs}^{hf}=\pi b_{c}^{ 2}$ at high frequencies, represented by the lines in Fig.~\ref{abstotald4d6}.}

\begin{figure}[!htb]
 \centering
 \subfigure[]{\includegraphics[scale=0.40]{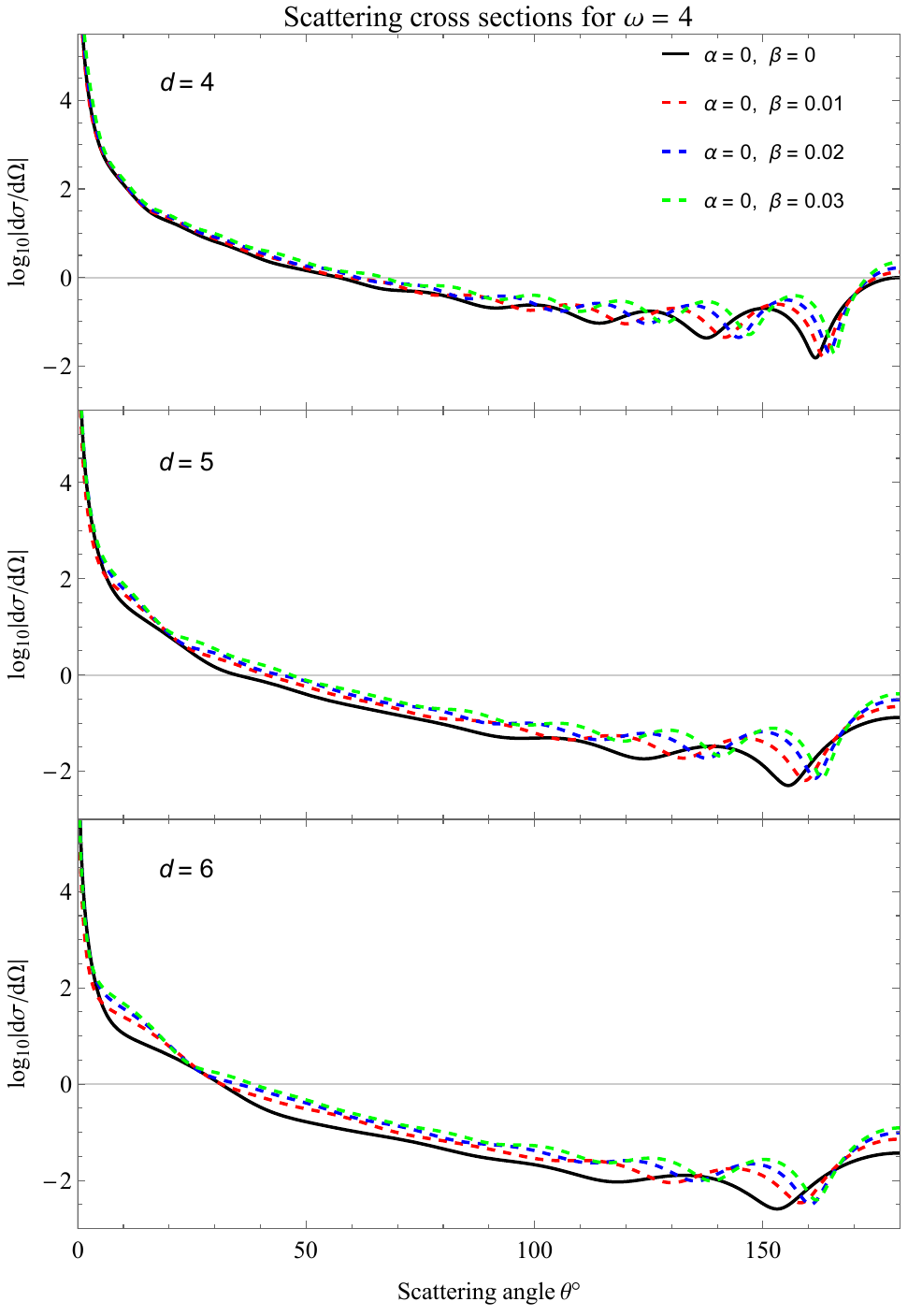}\label{scattd456a0}}
 \qquad
 \subfigure[]{\includegraphics[scale=0.40]{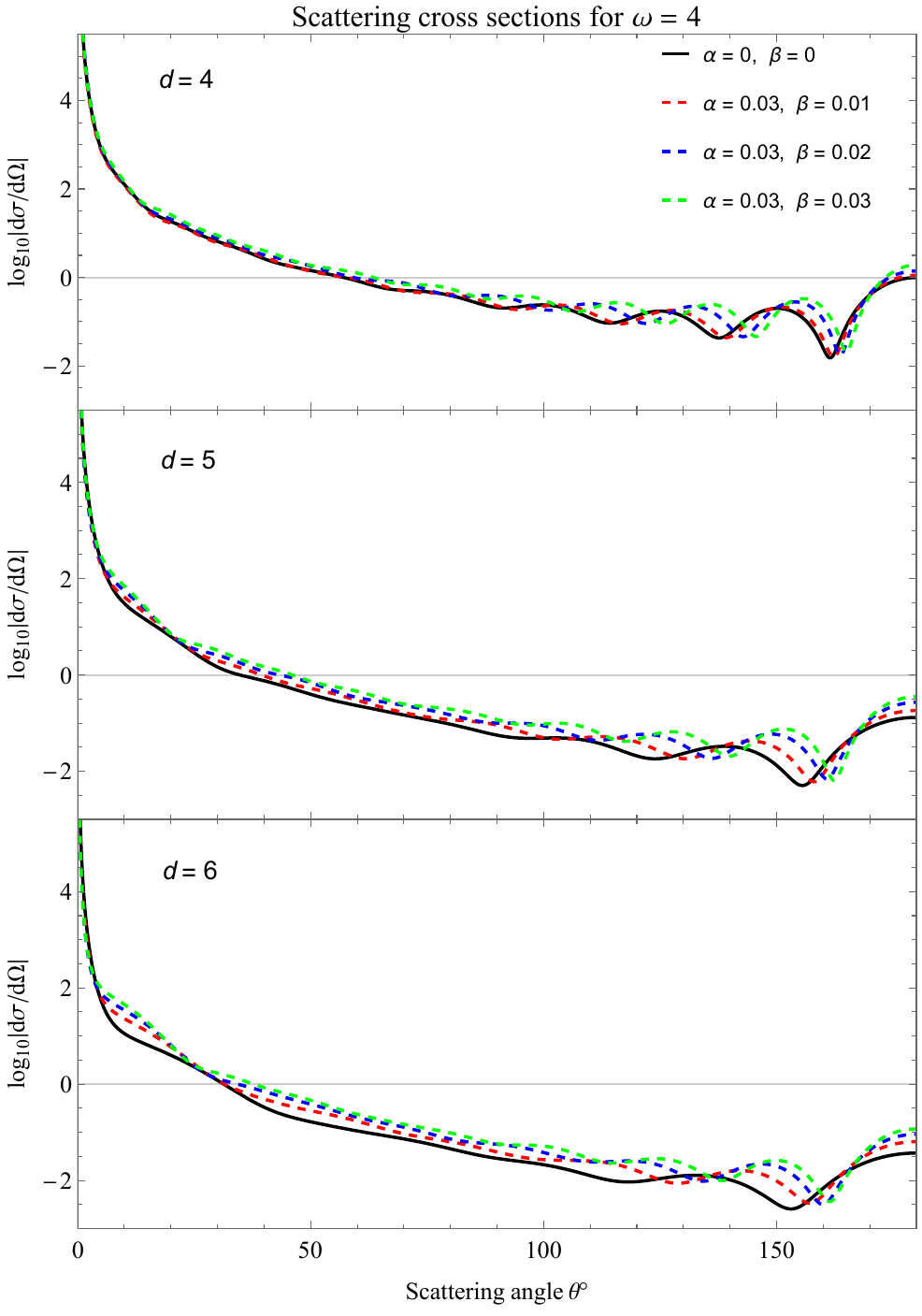}\label{scattd456a003}}
 \caption{\footnotesize{Differential scattering cross section for dimensions $d = 4, 5$ and $6$ at a frequency $\omega = 4$. We compare the conventional case $\alpha = 0$ and $\beta = 0$ with those corrected by GUP.}} 
 \label{scatt}
\end{figure}

\begin{figure}[!htb]
 \centering
 \subfigure[]{\includegraphics[scale=0.35]{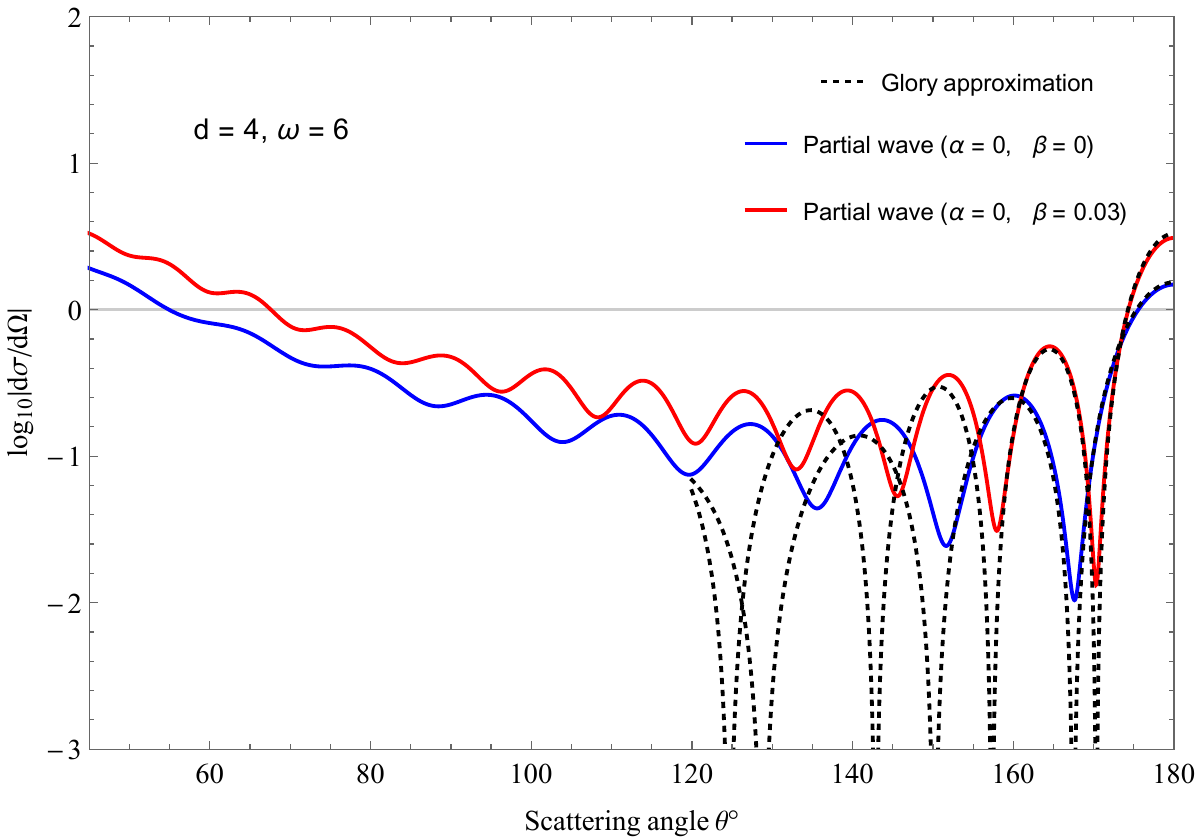}\label{gloryd4}}
 \qquad
 \subfigure[]{\includegraphics[scale=0.35]{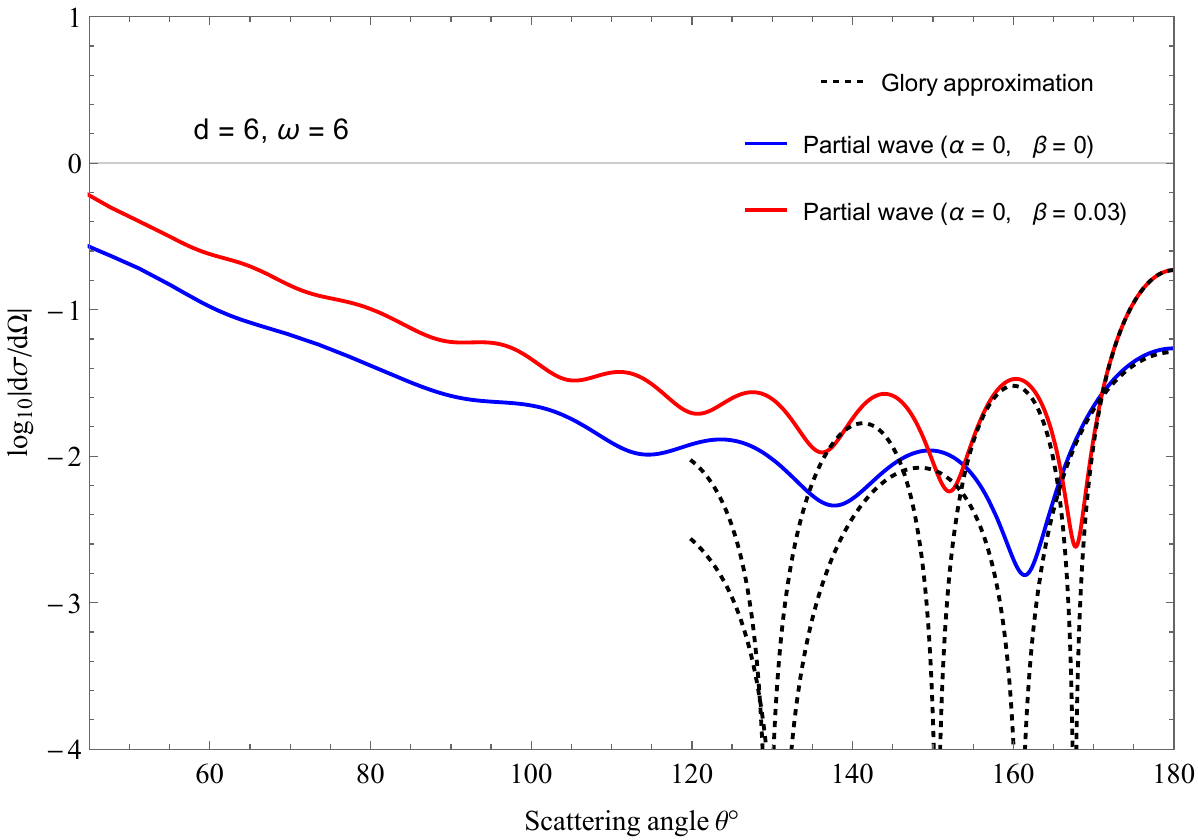}\label{gloryd6}}
 \\
  \caption{\footnotesize{Comparison for differential scattering cross section between partial waves and glory  approximation methods. Notice the behavior for dimensions $d=4$ and $d=6$ with the influence of the GUP correction term $\beta$. }}
 \label{glory}
\end{figure}

{Numerical results for the differential scattering cross section are shown in Figs.~\ref{scatt} and \ref{glory}. In equation \eqref{scattClass} and Fig.~\ref{ClassPlot}, we study the differential scattering cross section in the limit of small angles using classical analysis. However, the partial wave method allows us to verify the behavior in a complete interval of the angle of scattering, mainly at angles close to $180^{\circ}$. In Fig.~\ref{scatt}, we have the differential scattering cross section considering a frequency $\omega = 4$ for dimensions $d = 4, 5$, and $6$. See that the effect of the corrections coming from the GUP causes a change in the scattering curves, where it is smoother for $d=4$ mainly at small scattering angles, as also seen in Fig.~\ref{ClassPlot}. In Fig.~\ref{scattd456a0} we have only the contribution of the quadratic part $\beta$, while in Fig.~\ref{scattd456a003}, we have the two contributions $\alpha$ and $\beta$, but as we can see, we have no significant changes in the differential scattering cross section.}

{Fig.~\ref{glory} shows the comparison of the results for the differential scattering cross section using the partial wave method and semi-classical glory approximations using the values in Table~\ref{tab1} by fixing the value of the frequency at $\omega=6$. As we can see, the glory approximations adjust for large angles ($\theta \geq 160^{\circ}$). In Fig.~\ref{glory}, we can see in more detail the influence of the GUP correction parameter for these two dimensions, where, at higher angles, we have an increase in intensity and a shift in the interference fringes.}

\section{Conclusions}
\label{conc}
In this paper, we have carried out the process of scattering scalar waves by the Schwarzschild-Tangherlini black hole. 
We have found the phase shift analytically at the low and high frequency limits. 
We have analyzed the behavior of differential scattering/absorption cross-sections at the low and high frequency limits. 
Then, we have shown that the dominant contribution at the small angle limit of the differential scattering cross section is modified due to the contribution of the extra dimension.
We show that, in the limit of the mass parameter tending to zero, the absorption in high dimensions does not vanish. 
Besides, we have confirmed these results by numerically solving the radial equation for arbitrary frequencies.
{We have investigated the effect of quantum corrections implemented by GUP on the differential scattering cross section in different regimes. In the limit of small angles, in high dimensions, a reduction in the scattering cross section is seen in equation~\eqref{scattClass} and in Fig.~\ref{ClassPlot}. In these dimensions, we also have that the influence of the GUP corrections, mainly quadratic part $\beta $, is more present causing a small increase in the scattering cross section in each dimension. In addition, the effect of correction due to GUP in each dimension is more significant for large angles, as seen in Figs.~\ref{scatt} and \ref{glory}. In Fig.~\ref{glory}, we calculated the glory scattering by means of the semi-classical approximation~\eqref{scatglory}, and we saw that the results are compatible with the partial wave method.}

\acknowledgments

We would like to thank CNPq, CAPES and CNPq/PRONEX/FAPESQ-PB (Grant nos. 165/2018 and 015/2019),
for partial financial support. MAA, FAB and EP acknowledge support from CNPq (Grant nos. 306398/2021-4, 309092/2022-1, 304290/2020-3).


\begin{thebibliography}{100}





\bibitem{LIGOScientific:2016aoc}
B.~P.~Abbott \textit{et al.} [LIGO Scientific and Virgo],
``{\it Observation of Gravitational Waves from a Binary Black Hole Merger},''
Phys. Rev. Lett. \textbf{116}, no.6, 061102 (2016)
doi:10.1103/PhysRevLett.116.061102
[arXiv:1602.03837 [gr-qc]].

\bibitem{LIGOScientific:2017vwq}
B.~P.~Abbott \textit{et al.} [LIGO Scientific and Virgo],
``{\it GW170817: Observation of Gravitational Waves from a Binary Neutron Star Inspiral},''
Phys. Rev. Lett. \textbf{119}, no.16, 161101 (2017)
doi:10.1103/PhysRevLett.119.161101
[arXiv:1710.05832 [gr-qc]].

\bibitem{EventHorizonTelescope:2019dse}
K.~Akiyama \textit{et al.} [Event Horizon Telescope],
``{\it First M87 Event Horizon Telescope Results. I. The Shadow of the Supermassive Black Hole},''
Astrophys. J. Lett. \textbf{875}, L1 (2019)
doi:10.3847/2041-8213/ab0ec7
[arXiv:1906.11238 [astro-ph.GA]].

\bibitem{EventHorizonTelescope:2019ggy}
K.~Akiyama \textit{et al.} [Event Horizon Telescope],
``{\it First M87 Event Horizon Telescope Results. VI. The Shadow and Mass of the Central Black Hole},''
Astrophys. J. Lett. \textbf{875}, no.1, L6 (2019)
doi:10.3847/2041-8213/ab1141
[arXiv:1906.11243 [astro-ph.GA]].

\bibitem{Tangherlini:1963bw}
F.~R.~Tangherlini,
``{\it Schwarzschild field in n dimensions and the dimensionality of space problem},''
Nuovo Cim. \textbf{27}, 636-651 (1963)
doi:10.1007/BF02784569

\bibitem{Feng:2015jlj}
Z.~W.~Feng, H.~L.~Li, X.~T.~Zu and S.~Z.~Yang,
``{\it Corrections to the thermodynamics of Schwarzschild-Tangherlini black hole and the generalized uncertainty principle},''
Eur. Phys. J. C \textbf{76}, no.4, 212 (2016)
doi:10.1140/epjc/s10052-016-4057-1
[arXiv:1604.04702 [hep-th]].

\bibitem{Matzner1968} R. A. Matzner, ``{\it Scattering of Massless Scalar Waves by a Schwarzschild ``Singularity''} ",  J. Math. Phys. {\bf 9}, 163 (1968)
doi:10.1063/1.1664470

\bibitem{Futterman1988} J. A. Futterman, F. A. Handler, and R. A. Matzner,
{\it Scattering from black holes} (Cambridge University Press, England, 1988)
  

\bibitem{Matzner:1977dn}
R.~A.~Matzner and M.~P.~Ryan,
``{\it Low Frequency Limit of Gravitational Scattering},''
Phys. Rev. D \textbf{16}, 1636-1642 (1977)
doi:10.1103/PhysRevD.16.1636

\bibitem{Westervelt:1971pm}
P.~J.~Westervelt,
``{\it Scattering of electromagnetic and gravitational waves by a static gravitational field - comparison between the classical (general-relativistic) and quantum field-theoretic results},''
Phys. Rev. D \textbf{3}, 2319-2324 (1971)
doi:10.1103/PhysRevD.3.2319


\bibitem{Peters:1976jx}
P.~C.~Peters,
``{\it Differential Cross-Sections for Weak Field Gravitational Scattering},''
Phys. Rev. D \textbf{13}, 775-777 (1976)
doi:10.1103/PhysRevD.13.775


\bibitem{Sanchez:1976fcl}
N.~G.~Sanchez,
``{\it Scattering of scalar waves from a Schwarzschild black hole},''
J. Math. Phys. \textbf{17}, no.5, 688 (1976)
doi:10.1063/1.522949

\bibitem{Sanchez:1976xm}
N.~G.~Sanchez,
``{\it The Wave Scattering Theory and the Absorption Problem for a Black Hole},''
Phys. Rev. D \textbf{16}, 937-945 (1977)
doi:10.1103/PhysRevD.16.937


\bibitem{DeLogi:1977dp}
W.~K.~De Logi and S.~J.~Kovacs,
``{\it Gravitational Scattering of Zero Rest Mass Plane Waves},''
Phys. Rev. D \textbf{16}, 237-244 (1977)
doi:10.1103/PhysRevD.16.237


\bibitem{Doran:2001ag}
C.~Doran and A.~Lasenby,
``{\it Perturbation theory calculation of the black hole elastic scattering cross-section},''
Phys. Rev. D \textbf{66}, 024006 (2002)
doi:10.1103/PhysRevD.66.024006
[arXiv:gr-qc/0106039 [gr-qc]].


\bibitem{Dolan:2007ut} 
S. R. Dolan,``{\it Scattering of long-wavelength gravitational waves},'' Phys.\ Rev.\ D {\bf 77}, 044004 (2008) doi:10.1103/PhysRevD.77.044004 [arXiv:0710.4252 [gr-qc]].

\bibitem{Crispino:2009ki} 
  L.~C.~B.~Crispino, S.~R.~Dolan and E.~S.~Oliveira,
  ``{\it Scattering of massless scalar waves by Reissner-Nordstrom black holes},''
  Phys.\ Rev.\ D {\bf 79}, 064022 (2009)
  doi:10.1103/PhysRevD.79.064022
  [arXiv:0904.0999 [gr-qc]].


\bibitem{Churilov1974} A. A. Starobinsky and S. M. Churilov, Sov. Phys.- JETP {\bf 38}, 1 (1974).

\bibitem{Gibbons:1975kk}
G.~W.~Gibbons,
``{\it Vacuum Polarization and the Spontaneous Loss of Charge by Black Holes},''
Commun. Math. Phys. \textbf{44}, 245-264 (1975)
doi:10.1007/BF01609829


\bibitem{Page:1976df}
D.~N.~Page,
``{\it Particle Emission Rates from a Black Hole: Massless Particles from an Uncharged, Nonrotating Hole},''
Phys. Rev. D \textbf{13}, 198-206 (1976)
doi:10.1103/PhysRevD.13.198


\bibitem{Churilov1973}  A. A. Starobinskii and S. M. Churilov, Zh. Eksp. Teor. Fiz. {\bf 65}, 3 (1973).


\bibitem{Jung:2004yh}
E.~Jung and D.~K.~Park,
``{\it Effect of scalar mass in the absorption and emission spectra of Schwarzschild black hole},''
Class. Quant. Grav. \textbf{21}, 3717-3732 (2004)
doi:10.1088/0264-9381/21/15/007
[arXiv:hep-th/0403251 [hep-th]].


\bibitem{Doran2005} C. Doran, A. Lasenby, S. Dolan, and I. Hinder, Phys. Rev. D {\bf 71}, 124020 (2005), 
arXiv:gr-qc/0503019 [gr-qc].

\bibitem {Dolanprd2006} S. Dolan, C. Doran, and A. Lasenby, Phys. Rev. D {\bf 74}, 064005 (2006), arXiv:gr-qc/0605031 [gr-qc].

\bibitem{Castineiras2007} J. Castineiras, L. C. Crispino, and D. P. M. Filho, Phys. Rev. D {\bf 75}, 024012 (2007).
 
\bibitem{Benone:2014qaa} 
  C.~L.~Benone, E.~S.~de Oliveira, S.~R.~Dolan and L.~C.~B.~Crispino,
  ``{\it Absorption of a massive scalar field by a charged black hole},''
  Phys.\ Rev.\ D {\bf 89}, no. 10, 104053 (2014)
  doi:10.1103/PhysRevD.89.104053
  [arXiv:1404.0687 [gr-qc]].
  


\bibitem{Das:1996we} 
  S.~R.~Das, G.~W.~Gibbons and S.~D.~Mathur,
  ``{\it Universality of low-energy absorption cross-sections for black holes},''
  Phys.\ Rev.\ Lett.\  {\bf 78}, 417 (1997)
  doi:10.1103/PhysRevLett.78.417
  [hep-th/9609052]. 
     

\bibitem{deOliveira:2018kcq} 
  E.~S.~de Oliveira,
  ``{\it Scalar scattering from black holes with tidal charge},''
  Eur.\ Phys.\ J.\ C {\bf 78}, no. 11, 876 (2018)
  doi:10.1140/epjc/s10052-018-6316-9
  [arXiv:1805.04987 [gr-qc]].
  
\bibitem{Hai:2013ara} 
  H.~Hai, W.~Yong-Jiu and C.~Ju-Hua,
  ``{\it Absorption cross section of black holes with global monopole},''
  Chin.\ Phys.\ B {\bf 22}, no. 7, 070401 (2013). 
  

\bibitem{Sanchez1978}
N. G. S\'anchez,
``{\it Absorption and Emission Spectra of a Schwarzschild Black Hole},''
Phys. Rev. D \textbf{18}, 1030 (1978)
doi:10.1103/PhysRevD.18.1030


\bibitem{NSanchez1978}
N. G. S\'anchez, Rev. D {\bf 18}, 1798 (1978).



\bibitem{Campos:2021sff}
J.~A.~V.~Campos, M.~A.~Anacleto, F.~A.~Brito and E.~Passos,
``{\it Quasinormal modes and shadow of noncommutative black hole},''
Sci. Rep. \textbf{12}, no.1, 8516 (2022)
doi:10.1038/s41598-022-12343-w
[arXiv:2103.10659 [hep-th]].

\bibitem{Anacleto:2021qoe}
M.~A.~Anacleto, J.~A.~V.~Campos, F.~A.~Brito and E.~Passos,
``{\it Quasinormal modes and shadow of a Schwarzschild black hole with GUP},''
Annals Phys. \textbf{434}, 168662 (2021)
doi:10.1016/j.aop.2021.168662
[arXiv:2108.04998 [gr-qc]].

\bibitem{Anacleto:2018acl}
M.~A.~Anacleto, F.~A.~Brito, J.~A.~V.~Campos and E.~Passos,
``{\it Higher-derivative analogue Aharonov\textendash{}Bohm effect, absorption and superresonance},''
Int. J. Mod. Phys. A \textbf{35}, no.21, 2050112 (2020)
doi:10.1142/S0217751X20501122
[arXiv:1810.13356 [hep-th]].


\bibitem{Li:2022wzi}
Q.~Li, C.~Ma, Y.~Zhang, Z.~W.~Lin and P.~F.~Duan,
``{\it Shadow, absorption and Hawking radiation of a Schwarzschild black hole surrounded by a cloud of strings in Rastall gravity},''
Eur. Phys. J. C \textbf{82}, no.7, 658 (2022)
doi:10.1140/epjc/s10052-022-10623-3 

\bibitem{Xing:2022emg}
Y.~Xing, Y.~Yang, D.~Liu, Z.~W.~Long and Z.~Xu,
``{\it The ringing of quantum corrected Schwarzschild black hole with GUP},''
Commun. Theor. Phys. \textbf{74}, no.8, 085404 (2022)
doi:10.1088/1572-9494/ac7cdc
[arXiv:2204.11262 [gr-qc]].

\bibitem{Bisnovatyi-Kogan:2022ujt}
G.~S.~Bisnovatyi-Kogan and O.~Y.~Tsupko,
``{\it Analytical study of higher-order ring images of the accretion disk around a black hole},''
Phys. Rev. D \textbf{105}, no.6, 064040 (2022)
doi:10.1103/PhysRevD.105.064040
[arXiv:2201.01716 [gr-qc]].




\bibitem{Zeng:2021dlj}
X.~X.~Zeng, G.~P.~Li and K.~J.~He,
``{\it The shadows and observational appearance of a noncommutative black hole surrounded by various profiles of accretions},''
Nucl. Phys. B \textbf{974}, 115639 (2022)
doi:10.1016/j.nuclphysb.2021.115639
[arXiv:2106.14478 [hep-th]].


\bibitem{Mourad:2021qgo}
M.~F.~Mourad and M.~Abdelgaber,
``{\it Spherically symmetric AdS black holes with smeared mass distribution},''
Mod. Phys. Lett. A \textbf{36}, no.05, 2150029 (2021)
doi:10.1142/S0217732321500292

\bibitem{Jha:2021bue}
S.~K.~Jha and A.~Rahaman,
``{\it Lorentz violation and noncommutative effect on superradiance scattering off Kerr-like black hole and on the shadow of it},''
[arXiv:2111.02817 [gr-qc]].

\bibitem{Li:2022jda}
Q.~Li, C.~Ma, Y.~Zhang, Z.~W.~Lin and P.~F.~Duan,
``{\it Gray-body factor and absorption of the Dirac field in ESTGB gravity},''
Chin. J. Phys. \textbf{77}, 1269-1277 (2022)
doi:10.1016/j.cjph.2022.03.027

\bibitem{Gogoi:2022wyv}
D.~J.~Gogoi and U.~D.~Goswami,
``{\it Quasinormal modes and Hawking radiation sparsity of GUP corrected black holes in bumblebee gravity with topological defects},''
JCAP \textbf{06}, no.06, 029 (2022)
doi:10.1088/1475-7516/2022/06/029
[arXiv:2203.07594 [gr-qc]].

\bibitem{Karmakar:2022idu}
R.~Karmakar, D.~J.~Gogoi and U.~D.~Goswami,
``{\it Quasinormal modes and thermodynamic properties of GUP-corrected Schwarzschild black hole surrounded by quintessence},''
[arXiv:2206.09081 [gr-qc]].

\bibitem{Lobos:2022jsz}
N.~J.~L.~S.~Lobos and R.~C.~Pantig,
``{\it Generalized Extended Uncertainty Principle Black Holes: Shadow and lensing in the macro- and microscopic realms},''
[arXiv:2208.00618 [gr-qc]].

\bibitem{Tsupko:2022yzg}
O.~Y.~Tsupko,
``{\it Shape of higher-order photon rings: analytical description with polar curve},''
[arXiv:2208.02084 [gr-qc]].

\bibitem{Zeng:2022fdm}
X.~X.~Zeng, M.~I.~Aslam and R.~Saleem,
``{\it The Optical Appearance of Charged Four-Dimensional Gauss-Bonnet Black Hole with Strings Cloud and Non-Commutative Geometry Surrounded by Various Accretions Profiles},''
[arXiv:2208.06246 [gr-qc]].

\bibitem{Heydari-Fard:2021qdc}
M.~Heydari-Fard and M.~Heydari-Fard,
``{\it Null geodesics and shadow of 4D Einstein\textendash{}Gauss\textendash{}Bonnet black holes surrounded by quintessence},''
Int. J. Mod. Phys. D \textbf{31}, no.08, 2250066 (2022)
doi:10.1142/S0218271822500663
[arXiv:2109.02059 [gr-qc]].


\bibitem{Heydari-Fard:2021pjc}
M.~Heydari-Fard, M.~Heydari-Fard and H.~R.~Sepangi,
``{\it Null geodesics and shadow of hairy black holes in Einstein-Maxwell-dilaton gravity},''
Phys. Rev. D \textbf{105}, no.12, 124009 (2022)
doi:10.1103/PhysRevD.105.124009
[arXiv:2110.02713 [gr-qc]].


\bibitem{Khodadi:2021gbc}
M.~Khodadi, G.~Lambiase and D.~F.~Mota,
``{\it No-hair theorem in the wake of Event Horizon Telescope},''
JCAP \textbf{09}, 028 (2021)
doi:10.1088/1475-7516/2021/09/028
[arXiv:2107.00834 [gr-qc]].

\bibitem{Fathi:2020sfw}
M.~Fathi, M.~Olivares and J.~R.~Villanueva,
``{\it Gravitational Rutherford scattering of electrically charged particles from a charged Weyl black hole},''
Eur. Phys. J. Plus \textbf{136}, no.4, 420 (2021)
doi:10.1140/epjp/s13360-021-01441-9
[arXiv:2009.03404 [gr-qc]].

\bibitem{Chen:2022ngd}
H.~Chen, H.~Hassanabadi, B.~C.~L\"utf\"uo\u{g}lu and Z.~W.~Long,
``{\it Quantum corrections to the quasinormal modes of the Schwarzschild black hole},''
[arXiv:2203.03464 [gr-qc]].

\bibitem{Filho:2023etf}
A.~A.~A.~Filho, H.~Hassanabadi, N.~Heidari, J.~Kr\'\i{}z, P.~J.~Porf\'\i{}rio and S.~Zare,
``{\it Gravitational traces of bumblebee gravity in metric-affine formalism},''
[arXiv:2305.18871 [gr-qc]].

\bibitem{Heidari:2023bww}
N.~Heidari, H.~Hassanabadi, A.~A.~A.~Filho, J.~Kur\'\i{}uz, S.~Zare and P.~J.~Porf\'\i{}rio,
``{\it Gravitational signatures of a non--commutative stable black hole},''
[arXiv:2305.06838 [gr-qc]].


\bibitem{Myers:1986un}
R.~C.~Myers and M.~J.~Perry,
``{\it Black Holes in Higher Dimensional Space-Times},''
Annals Phys. \textbf{172}, 304 (1986)
doi:10.1016/0003-4916(86)90186-7

\bibitem{Jung:2004yn}
E.~Jung, S.~Kim and D.~K.~Park,
``{\it Proof of universality for the absorption of massive scalar by the higher-dimensional Reissner-Nordstrom black holes},''
Phys. Lett. B \textbf{602}, 105-111 (2004)
doi:10.1016/j.physletb.2004.09.067
[arXiv:hep-th/0409145 [hep-th]].

\bibitem{Moura:2011rr} 
  F.~Moura,
  ``{\it Scattering of spherically symmetric $d$-dimensional $\alpha'-$corrected black holes in string theory},''
  JHEP {\bf 1309}, 038 (2013)
  doi:10.1007/JHEP09(2013)038
  [arXiv:1105.5074 [hep-th]].

\bibitem{Marinho:2016ixt} 
  C.~I.~S.~Marinho and E.~S.~de Oliveira,
  ``{\it Scattering of massless scalar waves from Schwarzschild-Tangherlini black holes on the brane},''
  arXiv:1612.05604 [gr-qc].

\bibitem{Ahmedov:2021ohg}
B.~Ahmedov, O.~Rahimov and B.~Toshmatov,
``{\it Gravitational Capture Cross-Section of Particles by Schwarzschild-Tangherlini Black Holes},''
Universe \textbf{7}, no.8, 307 (2021)
doi:10.3390/universe7080307

\bibitem{Singh:2017vfr}
B.~P.~Singh and S.~G.~Ghosh,
``{\it Shadow of Schwarzschild\textendash{}Tangherlini black holes},''
Annals Phys. \textbf{395}, 127-137 (2018)
doi:10.1016/j.aop.2018.05.010
[arXiv:1707.07125 [gr-qc]].

\bibitem{Moura:2021eln}
F.~Moura and J.~Rodrigues,
``{\it Eikonal quasinormal modes and shadow of string-corrected d-dimensional black holes},''
Phys. Lett. B \textbf{819}, 136407 (2021)
doi:10.1016/j.physletb.2021.136407
[arXiv:2103.09302 [hep-th]].

\bibitem{Tsukamoto:2014dta}
N.~Tsukamoto, T.~Kitamura, K.~Nakajima and H.~Asada,
Phys. Rev. D \textbf{90}, no.6, 064043 (2014)
doi:10.1103/PhysRevD.90.064043
[arXiv:1402.6823 [gr-qc]].

\bibitem{das2008universality}
S.~Das and E.~C.~Vegenas,
``{\it Universality of quantum gravity corrections},''
Phys.Rev.Lett \textbf{101}, 221301 (2008)
doi.org/10.1103/PhysRevLett.101.221301
[arXiv:0810.5333v2 [hep-th]].

\bibitem{das2009phenomenological}
S.~Das and E.~C.~Vegenas,
``{\it Phenomenological implications of the generalized uncertainty principle},''
Can. J. Phys \textbf{87}, 233--240 (2009)
doi.org/10.1139/P08-105
[arXiv:0901.1768v1 [hep-th]].

\bibitem{ali2011proposal}
A.~F.~Ali, S.~Das and E.~C.~Vegenas,
``{\it Proposal for testing quantum gravity in the lab},''
Phys. Rev. D \textbf{84}, 044013 (2011)
doi.org/10.1103/PhysRevD.84.044013
[arXiv:1107.3164v2 [hep-th] ].

\bibitem{buoninfante2020phenomenology}
L.~Buoninfante, G.~Lambiase, G.~G.~Luciano and L.~Petruzziello,
``{\it Phenomenology of GUP stars},''
Eur. Phys. J. C \textbf{80}, 1--8 (2020)
doi:10.1140/epjc/s10052-020-08436-3
[arXiv:2001.05825 [gr-qc]].

\bibitem{Lake:2018zeg}
M.~J.~Lake, M.~Miller, R.~F.~Ganardi, Z.~Liu, S.~D.~Liang and T.~Paterek,
``{\it Generalised uncertainty relations from superpositions of geometries},''
Class. Quant. Grav. \textbf{36}, no.15, 155012 (2019)
doi:10.1088/1361-6382/ab2160
[arXiv:1812.10045 [quant-ph]].

\bibitem{Lake:2019nmn}
M.~J.~Lake, M.~Miller and S.~D.~Liang,
``{\it Generalised uncertainty relations for angular momentum and spin in quantum geometry},''
Universe \textbf{6}, no.4, 56 (2020)
doi:10.3390/universe6040056
[arXiv:1912.07094 [gr-qc]].

\bibitem{Casadio:2020rsj}
R.~Casadio and F.~Scardigli,
``\textit{Generalized Uncertainty Principle, Classical Mechanics, and General Relativity},''
Phys. Lett. B \textbf{807}, 135558 (2020)
doi:10.1016/j.physletb.2020.135558
[arXiv:2004.04076 [gr-qc]].

\bibitem{Buoninfante:2020guu}
L.~Buoninfante, G.~G.~Luciano, L.~Petruzziello and F.~Scardigli,
``\textit{Bekenstein bound and uncertainty relations},''
Phys. Lett. B \textbf{824}, 136818 (2022)
doi:10.1016/j.physletb.2021.136818
[arXiv:2009.12530 [hep-th]].

\bibitem{Anacleto:2020lel}
M.~A.~Anacleto, F.~A.~Brito, J.~A.~V.~Campos and E.~Passos,
``{\it Quantum-corrected scattering and absorption of a Schwarzschild black hole with GUP},''
Phys. Lett. B \textbf{810}, 135830 (2020)
doi:10.1016/j.physletb.2020.135830
[arXiv:2003.13464 [gr-qc]].

\bibitem{Koppel:2017rsf}
S.~K\"oppel, M.~Knipfer, M.~Isi, J.~Mureika and P.~Nicolini,
``\textit{Generalized uncertainty principle and extra dimensions},''
Springer Proc. Phys. \textbf{208}, 141-147 (2018)
doi:10.1007/978-3-319-94256-8\_16
[arXiv:1703.05222 [hep-th]].

\bibitem{Carr:2022ndy}
B.~J.~Carr,
``\textit{The Generalized Uncertainty Principle and higher dimensions: Linking black holes and elementary particles},''
Front. Astron. Space Sci. \textbf{9}, 1008221 (2022)
doi:10.3389/fspas.2022.1008221
[arXiv:2302.12609 [gr-qc]].

\bibitem{Scardigli:2008jn}
F.~Scardigli,
``\textit{Glimpses on the micro black hole Planck phase},''
Symmetry \textbf{12}, no.9, 1519 (2020)
doi:10.3390/sym12091519
[arXiv:0809.1832 [hep-th]].

\bibitem{Scardigli:2003kr}
F.~Scardigli and R.~Casadio,
{\it Generalized uncertainty principle, extra dimensions and holography},
Class. Quant. Grav. \textbf{20}, 3915-3926 (2003)
doi:10.1088/0264-9381/20/18/305
[arXiv:hep-th/0307174 [hep-th]].


\bibitem{Anacleto:2022shk}
M.~A.~Anacleto, F.~A.~Brito, J.~A.~V.~Campos and E.~Passos,
``{\it Absorption, scattering and shadow by a noncommutative black hole with global monopole},''
Eur. Phys. J. C \textbf{83}, no.4, 298 (2023)
doi:10.1140/epjc/s10052-023-11484-0
[arXiv:2212.13973 [hep-th]].

\bibitem{Anacleto:2017kmg}
M.~A.~Anacleto, F.~A.~Brito, S.~J.~S.~Ferreira and E.~Passos,
``{\it Absorption and scattering of a black hole with a global monopole in $f(R)$ gravity},''
Phys. Lett. B \textbf{788}, 231-237 (2019)
doi:10.1016/j.physletb.2018.11.020
[arXiv:1701.08147 [hep-th]].

\bibitem{Anacleto:2019tdj}
M.~A.~Anacleto, F.~A.~Brito, J.~A.~V.~Campos and E.~Passos,
``{\it Absorption and scattering of a noncommutative black hole},''
Phys. Lett. B \textbf{803}, 135334 (2020)
doi:10.1016/j.physletb.2020.135334
[arXiv:1907.13107 [hep-th]].

\bibitem{Anacleto:2020zhp}
M.~A.~Anacleto, F.~A.~Brito, J.~A.~V.~Campos and E.~Passos,
``{\it Absorption and scattering by a self-dual black hole},''
Gen. Rel. Grav. \textbf{52}, no.10, 100 (2020)
doi:10.1007/s10714-020-02756-1
[arXiv:2002.12090 [hep-th]].


\bibitem{medved2004conceptual}
A.~J.~M.~Medved and Elias~C.~Vagenas,
``{\it When conceptual worlds collide: the generalized uncertainty principle and the Bekenstein-Hawking entropy},''
Phys. Rev. D \textbf{70}, 124021 (2004)
doi.org/10.1103/PhysRevD.70.124021.
[arXiv:hep-th/0411022]

\bibitem{carr2015sub}
B.~Carr, J.~Mureika and P.~Nicolini,
``{\it Sub-Planckian black holes and the generalized uncertainty principle},''
JHEP \textbf{2015}, (1--24) (2015)
doi.org/10.48550/arXiv.1504.07637
[arXiv:1504.07637v2 [gr-qc]].

\bibitem{tawfik2014generalized}
A.~Tawfik and A.~Diab,
``{\it Generalized uncertainty principle: Approaches and applications},''
J. Mod. Phys. D \textbf{23}, 1430025 (2014)
doi.org/10.1142/S0218271814300250
[arXiv:1410.0206v2 [gr-qc]].

\bibitem{kempf1995hilbert}
A.~Tawfik and A.~Diab,
``{\it Hilbert space representation of the minimal length uncertainty relation},''
Phys. Rev. D. \textbf{52}, 1108 (1995)
doi.org/10.1103/PhysRevD.52.1108
[arXiv:hep-th/9412167v3].

\bibitem{Pedram:2011gw}
P.~Pedram,
``{\it A Higher Order GUP with Minimal Length Uncertainty and Maximal Momentum},''
Phys. Lett. B \textbf{714}, 317-323 (2012)
doi:10.1016/j.physletb.2012.07.005
[arXiv:1110.2999 [hep-th]].

\bibitem{park2008generalized}
M.~Park, 
``{\it The generalized uncertainty principle in (A) dS space and the modification of Hawking temperature from the minimal length},''
Phys. Rev. B. \textbf{659}, 698-702 (2008)
doi.org/10.1016/j.physletb.2007.11.090

\bibitem{Anacleto:2015mma}
M.~A.~Anacleto, F.~A.~Brito and E.~Passos,
``{\it Quantum-corrected self-dual black hole entropy in tunneling formalism with GUP},''
Phys. Lett. B \textbf{749}, 181-186 (2015)
doi:10.1016/j.physletb.2015.07.072
[arXiv:1504.06295 [hep-th]].

\bibitem{gangopadhyay2015constraints}
S.~Gangopadhyay, A.~Dutta and M.~Faizal
``{\it Constraints on the generalized uncertainty principle from black-hole thermodynamics},''
Euro. Phys. Lett. \textbf{112}, 20006 (2015)
doi.org/10.1209/0295-5075/112/20006
[arXiv:1501.01482v2 [gr-qc]]

\bibitem{Anacleto:2021nhm}
M.~A.~Anacleto, F.~A.~Brito, G.~C.~Luna and E.~Passos,
``{\it The generalized uncertainty principle effect in acoustic black holes},''
Annals Phys. \textbf{440}, 168837 (2022)
doi:10.1016/j.aop.2022.168837
[arXiv:2112.13573 [gr-qc]].

\bibitem{Segreto:2022clx}
S.~Segreto and G.~Montani,
``{\it Extended GUP formulation and the role of momentum cut-off},''
Eur. Phys. J. C \textbf{83}, no.5, 385 (2023)
doi:10.1140/epjc/s10052-023-11480-4
[arXiv:2208.03101 [quant-ph]].

\bibitem{feng2016quantum}
Z.~W.~Feng, H.~L.~Li and X~T~Zu, S.~Z.~Yang
``{\it Quantum corrections to the thermodynamics of Schwarzschild--Tangherlini black hole and the generalized uncertainty principle},''
Eur. Phys. J. C. \textbf{76}, 212 (2016)
doi.org/10.1140/epjc/s10052-016-4057-1
[arXiv:1512.09219v4 [hep-th]]

\bibitem{Feng:2016tyt}
Z.~W.~Feng, S.~Z.~Yang, H.~L.~Li and X.~T.~Zu,
``\textit{Constraining the generalized uncertainty principle with the gravitational wave event GW150914},''
Phys. Lett. B \textbf{768}, 81-85 (2017)
doi:10.1016/j.physletb.2017.02.043
[arXiv:1610.08549 [hep-ph]].

\bibitem{Feng:2020ams}
Z.~W.~Feng, G.~He, X.~Zhou, X.~L.~Mu and S.~Q.~Zhou,
``{\it Higher-order generalized uncertainty principle corrections to the Jeans mass},''
Eur. Phys. J. C \textbf{81}, no.8, 754 (2021)
doi:10.1140/epjc/s10052-021-09549-z
[arXiv:2006.01698 [physics.gen-ph]].

\bibitem{Feng:2022gdz}
Z.~W.~Feng, X.~Zhou and S.~Q.~Zhou,
``{\it Higher-order generalized uncertainty principle applied to gravitational baryogenesis},''
JCAP \textbf{06}, no.06, 022 (2022)
doi:10.1088/1475-7516/2022/06/022
[arXiv:2203.11671 [gr-qc]].

\bibitem{Anacleto:2022sim}
M.~A.~Anacleto, F.~A.~Brito, E.~Passos, J.~L.~Paulino, A.~T.~N.~Silva and J.~Spinelly,
``{\it Hawking radiation and entropy of a BTZ black hole with minimum length},''
Mod. Phys. Lett. A \textbf{37}, no.32, 2250215 (2022)
doi:10.1142/S0217732322502157
[arXiv:2301.05970 [hep-th]].

\bibitem{Scardigli:1999jh}
F.~Scardigli,
{\it Generalized uncertainty principle in quantum gravity from micro - black hole Gedanken experiment},
Phys. Lett. B \textbf{452}, 39-44 (1999)
doi:10.1016/S0370-2693(99)00167-7
[arXiv:hep-th/9904025 [hep-th]].

\bibitem{Scardigli:2014qka}
F.~Scardigli and R.~Casadio,
{\it Gravitational tests of the Generalized Uncertainty Principle},
Eur. Phys. J. C \textbf{75}, no.9, 425 (2015)
doi:10.1140/epjc/s10052-015-3635-y
[arXiv:1407.0113 [hep-th]].

\bibitem{Scardigli:2016pjs}
F.~Scardigli, G.~Lambiase and E.~Vagenas,
{\it GUP parameter from quantum corrections to the Newtonian potential},
Phys. Lett. B \textbf{767}, 242-246 (2017)
doi:10.1016/j.physletb.2017.01.054
[arXiv:1611.01469 [hep-th]].

\bibitem{Scardigli:2018jlm}
F.~Scardigli, M.~Blasone, G.~Luciano and R.~Casadio,
{\it Modified Unruh effect from Generalized Uncertainty Principle},
Eur. Phys. J. C \textbf{78}, no.9, 728 (2018)
doi:10.1140/epjc/s10052-018-6209-y
[arXiv:1804.05282 [hep-th]].

\bibitem{Lambiase:2017adh}
G.~Lambiase and F.~Scardigli,
``Lorentz violation and generalized uncertainty principle,''
Phys. Rev. D \textbf{97}, no.7, 075003 (2018)
doi:10.1103/PhysRevD.97.075003
[arXiv:1709.00637 [hep-th]].

\bibitem{adler2001generalized}
R.~J.~Adler, P.~Chen and P~Santiago
``{\it The generalized uncertainty principle and black hole remnants},''
Gen. Relativ. Gravit. \textbf{33}, 2101-2108 (2001)
doi.org/10.1023/A\% 3A1015281430411
[arXiv:gr-qc/0106080]

\bibitem{Scardigli:2010gm}
F.~Scardigli, C.~Gruber and P.~Chen,
``\textit{Black Hole Remnants in the Early Universe},''
Phys. Rev. D \textbf{83}, 063507 (2011)
doi:10.1103/PhysRevD.83.063507
[arXiv:1009.0882 [gr-qc]].

\bibitem{matzner1985glory}
R.~A.~Matzner, C.~DeWitte-Morette, B.~Nelson and T.~R.~Zhang,
``{\it Glory scattering by black holes},''
Phys. Rev. D. \textbf{31}, 1869 (1985)
doi.org/10.1103/PhysRevD.31.1869

\bibitem{crispino2009scattering}
L.~C.~B.~Crispino, S.~R.~Dolan and E.~S.~Oliveira,
``{\it Scattering of massless scalar waves by Reissner-Nordstr{\"o}m black holes},''
Phys. Rev. D. \textbf{79}, 064022 (2009)
doi.org/10.1103/PhysRevD.79.064022
[arXiv:0904.0999 [gr-qc]]

\bibitem{macedo2015scattering}
C.~F.~B.~Macedo, E.~S.~Oliveira and L.~C.~B.~Crispino,
``{\it Scattering by regular black holes: planar massless scalar waves impinging upon a Bardeen black hole},''
Phys. Rev. D. \textbf{92}, 024012 (2015)
doi.org/10.1103/PhysRevD.92.024012
[arXiv:1505.07014 [gr-qc]]

\bibitem{Yennie:1954zz}
D.~R.~Yennie, D.~G.~Ravenhall and R.~N.~Wilson,
``{\it Phase-Shift Calculation of High-Energy Electron Scattering},''
Phys. Rev. \textbf{95}, 500-512 (1954)
doi:10.1103/PhysRev.95.500

\bibitem{Cotaescu:2014jca}
I.~I.~Cotaescu, C.~Crucean and C.~A.~Sporea,
``{\it Partial wave analysis of the Dirac fermions scattered from Schwarzschild black holes},''
Eur. Phys. J. C \textbf{76}, no.3, 102 (2016)
doi:10.1140/epjc/s10052-016-3936-9
[arXiv:1409.7201 [gr-qc]].


\bibitem{Jung:2004nh}
E.~Jung, S.~Kim and D.~K.~Park,
``{\it Low-energy absorption cross section for massive scalar and Dirac fermion by (4+n)-dimensional Schwarzschild black hole},''
JHEP \textbf{09}, 005 (2004)
doi:10.1088/1126-6708/2004/09/005
[arXiv:hep-th/0406117 [hep-th]].

\bibitem{Dolan:2009zza}
S.~R.~Dolan, E.~S.~Oliveira and L.~C.~B.~Crispino,
``{\it Scattering of sound waves by a canonical acoustic hole},''
Phys. Rev. D \textbf{79}, 064014 (2009)
doi:10.1103/PhysRevD.79.064014
[arXiv:0904.0010 [gr-qc]].

\bibitem{Morse:1954} P. M. Morse and H. Feshbach, ``{\it Methods of theoretical physics}," American Journal of
Physics {\bf 22}, 410 (1954).

\bibitem{Ford:2000uye}
K.~W.~Ford and J.~A.~Wheeler,
``{\it Semiclassical Description of Scattering},''
Annals Phys. \textbf{281}, no.1-2, 608-635 (2000)
doi:10.1006/aphy.2000.6018


\bibitem{Decanini:2011xw}
Y.~Decanini, A.~Folacci and B.~Raffaelli,
``{\it Fine structure of high-energy absorption cross sections for black holes},''
Class. Quant. Grav. \textbf{28}, 175021 (2011)
doi:10.1088/0264-9381/28/17/175021
[arXiv:1104.3285 [gr-qc]].

\bibitem{dInverno} R. A. D'Inverno, ``{\it Introducing Einstein’s relativity}," Clarendon Press, 1992.

















\end{thebibliography}

\end{document}